\documentclass[12pt]{article}
\pdfoutput=1
\usepackage{epsf,amsfonts,amssymb,epsfig,amsmath,mathtools,graphics,slashed}
\usepackage{hyperref,hep,graphicx,subfig}

\newcommand{\nn}{\notag \\}

\newcommand{\hsce}{\zeta}

\makeatletter

\@addtoreset{equation}{section}
\makeatother

\begin{document}

\begin{titlepage}

\vfill

\begin{flushright}
Imperial/TP/2015/JG/03\\
DCPT-15/43
\end{flushright}

\vfill

\begin{center}
   \baselineskip=16pt
   {\Large\bf Thermoelectric DC conductivities and\\Stokes flows on
  black hole horizons }
  \vskip 1.5cm
  \vskip 1.5cm
      Elliot Banks$^1$, Aristomenis Donos$^2$ and Jerome P. Gauntlett$^1$\\
   \vskip .6cm
         \vskip .6cm
      \begin{small}
      \textit{$^1$Blackett Laboratory, 
        Imperial College\\ London, SW7 2AZ, U.K.}
        \end{small}\\
    \vskip .6cm
      \begin{small}
      \textit{$^2$Centre for Particle Theory\\ and Department of Mathematical Sciences\\Durham University\\Durham, DH1 3LE, U.K.}
        \end{small}\\
         
\end{center}

\vfill

\begin{center}
\textbf{Abstract}
\end{center}

\begin{quote}
We consider a general class of electrically charged black holes of Einstein-Maxwell-scalar theory that are holographically 
dual to conformal field theories at finite charge density which
break translation invariance explicitly. We examine the linearised perturbations about the solutions that are
associated with the thermoelectric DC conductivity. We show that there
is a decoupled sector at the black hole horizon which must solve generalised Stokes equations for a charged fluid. By solving these equations we can obtain the DC conductivity of the dual field theory. 
For Q-lattices and one-dimensional lattices we solve the fluid equations to obtain closed form expressions for the
DC conductivity in terms of the solution at the black hole horizon. We also determine the leading order
DC conductivity for lattices that can be expanded as a perturbative series about translationally invariant solutions.
\end{quote}

\vfill

\end{titlepage}

\setcounter{equation}{0}

\section{Introduction}
The holographic correspondence provides a powerful framework for obtaining precise results about strongly coupled systems using weakly coupled gravitational descriptions. A key cornerstone is that the dual description of the field theory at finite temperature is provided by a black hole spacetime. It is a remarkable fact that various properties of the thermal system are captured by the properties of the gravitational solutions at the black hole horizon. For example, the entropy of the thermal system at equilibrium is given by the Bekenstein-Hawking forumla, $S=A/4G$, where $A$ is the area of the black hole event horizon. Another well-known example is provided by
the shear viscosity, $\eta$, which, for certain classes of black hole solutions, is given by 
$\eta=4\pi s$, where $s$ is the entropy density \cite{Policastro:2001yc,Kovtun:2004de}.
In this paper, which expands upon and generalises the results presented in\cite{Donos:2015gia}, we explain how the DC thermo-electric conductivity
can be obtained by solving equations for a non-relativistic fluid on the black hole horizon. 
Moreover, we will see that, unlike $\eta$, our result for the DC conductivity holds in a very general context, being applicable to
arbitrary static black holes for which the DC conductivity is finite. The extension to stationary black holes will be presented in \cite{dggm}.

We first recall that the DC thermal conductivity, $\kappa$, is a very natural observable to study in holography. Indeed, in the regime of linear response it determines the heat current, or momentum flow, that is produced after applying a constant external thermal gradient. 
Although this naively appears to be a low-energy quantity it is in fact sensitive to the UV physics. For example, if the system is translationally invariant, and hence conserves momentum, then $\kappa$ is infinite. More precisely, there is a delta function in the AC conductivity at $\omega=0$. To obtain a finite $\kappa$ it is necessary to introduce a mechanism for momentum to dissipate.

For systems with a $U(1)$ symmetry, another natural
observable to consider is the DC electric conductivity. Momentum dissipation is also required in order to obtain a finite result when the 
charge density is non-vanishing. More generally, 
for these systems there is a mixing of electric and heat currents and one should consider the matrix of thermo-electric conductivities
\begin{align}\label{bigform}
\left(
\begin{array}{c}
\bar J\\\bar Q
\end{array}
\right)=
\left(\begin{array}{cc}
\sigma & \alpha T \\
\bar\alpha T &\bar\kappa T   \\
\end{array}\right)
\left(
\begin{array}{c}
E\\-(\nabla T)/T
\end{array}
\right)\,,
\end{align}
where $\bar J$ and $\bar Q$ 
are the total electric and heat current flux densities (which we define precisely later)
and $E$ and $\nabla T$ are constant applied electric field and thermal gradients, respectively.

Holographic lattices provide a natural framework\footnote{Another approach is to use massive gravity
and various interesting results have been obtained e.g. \cite{Vegh:2013sk,Blake:2013owa,Davison:2013txa,Amoretti:2014mma,Baggioli:2014roa,Amoretti:2015gna}. 
The DC conductivity has also been examined in the context of
translationally invariant probe branes e.g. \cite{Karch:2007pd,Hartnoll:2009ns}. In these constructions a
finite DC conductivity can arise because the delta function is suppressed by $1/N$, where $N$ is the number
of branes providing the background geometry.}
for studying momentum dissipation. These are black hole solutions 
whose asymptotic behaviour at the holographic boundary corresponds to
the addition of spatially dependent sources to the dual field theory. Various holographic lattices
with sources that depend periodically on just one of the non-compact spatial directions have been constructed
by solving PDEs in two variables 
\cite{Horowitz:2012ky,Horowitz:2012gs,Chesler:2013qla,Ling:2013nxa,Balasubramanian:2013yqa,Donos:2014yya,Rangamani:2015hka}. 
Constructing lattices that depend periodically
on additional spatial dimensions generically requires solving PDEs in more variables, which becomes increasingly challenging at the technical level. An important class of exceptions are provided by Q-lattices \cite{Donos:2013eha,Donos:2014uba}
which exploit a global symmetry in the bulk leading to a problem involving ODEs.
Other constructions involving ODEs use massless `axions' to obtain sources linear
in the spatial coordinates \cite{Azeyanagi:2009pr,Mateos:2011ix,Mateos:2011tv,Andrade:2013gsa,Cheng:2014qia,Jain:2014vka}, or use the metric or matter fields to obtain helical sources
when $D\ge 5$
\cite{Donos:2012js,Donos:2014gya,Donos:2014oha,Erdmenger:2015qqa}. A particularly
interesting application of holographic lattices is that they can lead to novel
incoherent metal ground states \cite{Donos:2012js,Donos:2014uba,Gouteraux:2014hca}, insulating ground states \cite{Donos:2012js,Donos:2013eha,Donos:2014uba} and transitions
between them \cite{Donos:2012js,Donos:2014uba}. Connections between
holographic lattices and superconductivity have also been explored in 
\cite{Ling:2014laa,Erdmenger:2015qqa,Andrade:2014xca}.

For the special case of translationally invariant black holes at zero chemical
potential, while the thermal conductivity is infinite the electric conductivity is finite and can be expressed in closed form in terms of the behaviour of the solution at the black hole horizon \cite{Iqbal:2008by}. 
For the above holographic lattices involving ODEs, formulae for the thermoelectric DC conductivity, also expressed in terms of the black hole solutions at the black hole horizon, were obtained in \cite{Donos:2014uba,Donos:2014cya,Donos:2014gya,Donos:2014oha}. These results were then extended to a class of one-dimensional holographic lattices in \cite{Donos:2014yya}, although the details were
more involved. Given these results it is natural to anticipate that similar results can be obtained for all holographic lattices. Here we show that for a broad class of 
holographic lattices, depending on all spatial directions, in general one cannot obtain such explicit formulae for
the DC conductivity. However, it is possible to
obtain the DC conductivity
after solving a set of fluid equations on the black hole horizon. These 
equations are a generalisation of the forced Stokes equations for a
charged fluid on the curved black hole horizon, with additional viscous terms
arising from bulk scalar fields. Recall that the Stokes equations are 
a time-independent and linearised limit of the
Navier-Stokes equations for an incompressible fluid arising at low Reynolds numbers. 
We will show that the previous results on the DC conductivity can all be obtained as special cases where the
Stokes equations can be solved explicitly in closed form.

The fact that the fluid equations which arise are linear and time-independent is not too surprising since
we are calculating in the regime of linear response and we are calculating the DC
conductivity. Similarly, the forcing terms are very natural since they arise from the applied sources for the
electric and heat currents. On the other hand only a subset of the linearised perturbation appears in the equations on the horizon and it is remarkable
that the equations form a closed system.

We emphasise that unlike in the relativistic fluid-gravity correspondence \cite{Bhattacharyya:2008jc}, and the associated
non-relativistic limit \cite{Bhattacharyya:2008kq,Fouxon:2008tb}, we do not take any hydrodynamic limit in obtaining our fluid equations. In the presence of spatially dependent sources
a natural hydrodynamic limit would arise for temperatures much bigger than
all other scales, including those of the lattice. By contrast our results are valid for all
temperatures. Our results differ but are also reminiscent 
of the ``membrane paradigm" \cite{Price:1986yy} and the more recent work\footnote{Ref.  \cite{Bredberg:2011jq} also contains a discussion and references to some of the earlier work on fluids and black hole horizons.} 
which relates solutions of the non-linear
Navier-Stokes equations on hypersurfaces in Minkowski space to obtain black hole solutions \cite{Bredberg:2011jq} (see also 
\cite{Compere:2011dx,Huang:2011he,Lysov:2013jsa}). We expect that the time-dependent and non-linear
versions of our equations will play a role in studying momentum dissipation for holographic lattices, possibly after taking a hydrodynamic limit and this
will be reported on elsewhere. Discussions of momentum dissipation, conductivities and hydrodynamics can be found in \cite{Hartnoll:2007ih,Davison:2014lua,Blake:2015epa,Lucas:2015lna} and some recent low-frequency conductivity results are presented in \cite{Davison:2015bea}.

The plan of the rest of the paper is as follows. In section 2 we introduce the holographic model and the class of
electrically charged black hole solutions that we shall be considering. In section 3 we analyse the linearised
perturbations, containing sources for the electric and heat currents, which are associated with the DC conductivity. We will show how the Stokes equations
can be obtained by expanding the Hamiltonian,
momentum and Gauss law constraints, associated with
a radial hamiltonian decomposition, at the black hole horizon. The fluid equations determine electric and heat currents
at the horizon in terms of sources for the electric and heat currents. In turn these can be used to obtain
suitably defined constant electric and heat current fluxes which are independent of the radial direction and hence give rise
to the DC conductivity. If the deformed CFT is living on $\Sigma_d$ then the DC conductivity is a
$b_1(\Sigma_d)\times b_1(\Sigma_d)$ matrix, where $b_1(\Sigma_d)$ is the first Betti number of $\Sigma_d$.
We emphasise that we provide a precise procedure for calculating the DC conductivity of the boundary theory
(the spectral weight of a two-point function) by solving an auxiliary set of fluid equations on the black hole horizon.

In section 4 we analyse some examples. We first generalise our results to an arbitrary number of
scalar fields which allows us to reconsider Q-lattices. For the Q-lattices and also for general one-dimensional lattices, we show that the fluid equations can be explicitly solved and we can obtain formulae for the DC conductivity explicitly in terms of the black hole solution at the horizon. 
We also examine holographic lattices that can be obtained as perturbative expansions about translationally
invariant solutions, including the AdS-RN black brane. We show that the leading order DC conductivity can also be found in closed form.
We briefly conclude in section 5 where we put some of the main results in a more general setting of general static
black hole spacetimes.
The paper contains four
appendices, including a discussion of the Hamiltonian decomposition of the equations of motion
with respect to the radial coordinate in appendix A.

\section{The background black holes}

We will consider theories in $D$ spacetime dimensions which couple the metric
to a gauge-field, $A$, and a single scalar field, $\phi$. The extension of our analysis
to additional scalar fields is straightforward as we will discuss later. 
We focus on $D\ge 4$.
The action is given by
\begin{align}\label{eq:bulk_action}
S=\int d^D x \sqrt{-g}\,\left(R-V(\phi)-\frac{Z(\phi)}{4}\,F^{2}-\frac{1}{2}\left(\partial\phi \right)^2\right)\,.
\end{align}
The equations of motion are given by
\begin{align}\label{geneoms}
R_{\mu\nu}-\frac{V}{D-2}g_{\mu\nu}-\frac{1}{2}\partial_\mu\phi\partial_\nu\phi-\frac{1}{2}Z(\phi)\left(
F_{\mu\rho}F_{\nu}{}^{\rho}- \frac{1}{2(D-2)}g_{\mu\nu}\,F^2\right)&=0\,,\nn
 \nabla_{\mu}\left[Z(\phi)F^{\mu\nu}\right]&=0\,,\nn
 \nabla^2\phi-V'(\phi)-\frac{1}{4}Z'(\phi)F^2&=0\,.
 \end{align}
 
The only restrictions that we will make on the functions $V(\phi)$, $Z(\phi)$ is that
$V(0)=-(D-1)(D-2)$, $V'(0)=0$ and $Z(0)=constant$. This ensures that a unit radius $AdS_{D}$ solves
the equations of motion with $\phi=0$ and this is dual to a CFT with a stress tensor,
dual to the metric, a 
global $U(1)$ current, dual to $A$, and an additional operator dual to $\phi$. 
When $Z'(0)=0$, the $D$-dimensional electrically charged AdS-Reissner-Nordstr\"om black hole also
solves the equations of motion and describes the CFT at constant charge
density. Note that we have set $16\pi G=1$, as well as setting
the AdS radius to unity, for convenience.
 
We will focus on a general class of electrically charged static black holes with metric and gauge-field
given by
 \begin{align}\label{eq:DC_ansatz}
ds^{2}&=-UG\,dt^{2}+\frac{F}{U}\,dr^{2}+ds^2(\Sigma_d)\,, \nn
A&=a_{t}\,dt\,,
\end{align}
where
$ds^2(\Sigma_d)\equiv g_{ij}(r,x)dx^i dx^j$ is a metric on a ($d\equiv D-2$)-dimensional manifold, $\Sigma_d$,
at fixed $r$. In addition, $U=U(r)$, while $G,F,a_t$ and $\phi$ are all functions of 
$(r,x^i)$. In section \ref{disc} we will discuss our main results in the context of a more general class of static black hole
solutions.

Asymptotically, as $r\to\infty$, the solutions are taken to approach $AdS_D$ with
\begin{align}\label{asymptsol}
&U\to r^2,\qquad F\to 1,\qquad 
G\to \bar G(x),\quad g_{ij}(r,x)\to r^2 \bar g_{ij}(x),\nn
&a_t(r,x)\to \mu(x),\qquad \phi(r,x)\to  r^{\Delta-d-1}\bar \phi(x)\,.
\end{align}
The spatial dependence of the boundary metric given by $\bar G(x)$,
$\bar g_{ij}(x)$ corresponds to providing
a source for the stress tensor in the dual CFT living on $\mathbb{R}\times\Sigma_d$. Similarly, $\mu(x)$ is a spatially dependent chemical potential for the
global abelian symmetry
and $\bar\phi(x)$ gives rise to a spatially dependent source for the associated
dual operator, which we have assumed has scaling
dimension $\Delta$.

A particularly interesting class of black holes is associated with adding sources to CFTs in flat Minkowski space, $\mathbb{R}^{1,d}$. In this case $\Sigma_d$ is topologically $\mathbb{R}^d$. 
Periodic lattices, which have been a focus of study, are obtained by taking the functions
$\bar G(x)$, $\bar g_{ij}(x)$, $\mu(x)$ and $\bar\phi(x)$
to be periodic functions on $\mathbb{R}^d$. If we denote the period in each of the spatial directions be $L_i$ then for this class of black holes we can, in effect, 
take $\Sigma_d$ to parametrise a torus with periods  $x^i\sim x^i+ L_i$.

The black hole horizon, which has topology $\Sigma_d$, is assumed to be located at $r=0$. 
By considering the Kruskal coordinate $v=t+\frac{\ln r}{4\pi T}+\dots$ we deduce that the
near horizon expansions are given by 
\begin{align}\label{nhexpbh}
U\left(r\right)&=r\left(4\pi\,T+U^{(1)}\,r+\dots\right)\,,\nn
a_{t}(r,x)&=r\left(a^{(0)}_{t}G^{(0)}\left(x\right)+a^{(1)}_{t}\left(x\right)r+\dots\right)\,,\nn
G(r,x)&=G^{(0)}\left(x\right)+G^{(1)}\left(x\right)r+\dots\,,\nn
F(r,x)&=F^{(0)}\left(x\right)+F^{(1)}\left(x\right)r+\dots\,,\nn
g_{ij}&=g_{ij}^{(0)}+g_{ij}^{(1)}r+\dots\,,\nn
\phi&=\phi^{(0)}(x)+r\phi^{(1)}(x)+\dots\,,
\end{align}
with
\begin{align}\label{gandf}
G^{(0)}\left(x\right)=F^{(0)}\left(x\right)\,.
\end{align}
We have added the extra factor of $G^{(0)}$ in the leading expression for $a_t$ for convenience. 
For later use, we observe that electric charge density at the horizon is simply
\begin{align}
\rho_H=\sqrt{-g}Z(\phi)F^{tr}|_H=\sqrt{-g_0}a^{(0)}_{t}Z^{(0)}\,,
\end{align}
where $Z^{(0)}\equiv Z(\phi^{(0)})$.
For the averaged holographic charge density, $\rho$, we have 
\begin{align}\label{rhotot}
\rho\equiv \frac{1}{\text{vol}_d}\int d^dx (\sqrt{-g}Z (\phi)F^{tr})|_\infty
=\frac{1}{\text{vol}_d}\int d^dx \rho_H\,,
\end{align}
where $\text{vol}_d\equiv \int d^dx \sqrt{-\bar g}$ is the volume of the
spatial metric at the $AdS$ boundary. This result follows from the fact that the
gauge-equations of motion implies $\partial_r(\sqrt{-g}ZF^{tr})+\partial_i(\sqrt{-g}ZF^{ti})=0$ and for the case of non-compact $\Sigma_d$ we have assumed any boundary terms vanish.

This class of black holes includes almost all of the holographic lattices that have been constructed to date as special cases.
For example, periodic lattices with modulated chemical potential with 
non-vanishing zero mode were studied in
\cite{Horowitz:2012ky,Ling:2013nxa,Donos:2014yya}, while the case of vanishing zero mode was studied in \cite{Chesler:2013qla}. Periodic lattices with a single
real scalar field have been studied in \cite{Horowitz:2012gs,Rangamani:2015hka}. These examples have spatially inhomogeneous sources in one direction only.
By contrast, the Q-lattice construction using two (or more) scalar fields \cite{Donos:2013eha,Donos:2014uba,Gouteraux:2014hca,Donos:2014cya,Kiritsis:2015oxa}
the non-periodic `axionic' lattices studied in \cite{Azeyanagi:2009pr,Mateos:2011ix,Mateos:2011tv,Andrade:2013gsa,Cheng:2014qia,Jain:2014vka,Donos:2014cya,Jain:2015txa} are homogeneous and the sources can be in any number of the spatial directions.
Other homogeneous constructions using $D=5$ helical lattices have been
studied in \cite{Donos:2014gya,Donos:2014oha} (an additional gauge field is needed to be included
to cover the examples of \cite{Donos:2012js}).
Metric deformations in one spatial dimension were studied in \cite{Balasubramanian:2013yqa}.
Holographic lattices in the presence of magnetic fields have been studied in \cite{Blake:2015ina}; the generalisation of our results to include magnetic fields
will be discussed elsewhere \cite{dggm}.

\section{Perturbing the black holes}
We want to study the holographic, linear response of the black holes
after applying suitable one-form sources $(E,\zeta)$ on $\Sigma_d$
for the electric and heat currents,
respectively.
Generalising \cite{Donos:2014uba,Donos:2014cya,Donos:2014yya} we incorporate the sources by the addition of terms that are linear in time. Specifically, we 
consider the following linear perturbation:
\begin{align}\label{pertansatz}
\delta \left(ds^2\right)&=\delta g_{\mu\nu} dx^\mu dx^\nu-2t M\zeta _i dt dx^i\,,\nn
\delta A&=\delta a_\mu dx^\mu-t E_i dx^i+t N \zeta_i dx^i\,,\nn
\delta\phi\,.
\end{align}
Here, $\delta g_{\mu\nu}$,$\delta a_\mu$, $\delta\phi$ are all functions of $(r,x^i)$,
while $E_i=E_i(x)$, $\zeta_i=\zeta_i(x)$.
We demand that $E,\zeta$ are closed one-forms on $\Sigma_d$:
\begin{align}\label{closed}
d(E_i dx^i)=d(\zeta _i dx^i)=0\,.
\end{align}
This means that the one forms $E,\zeta$ can uniquely be written as the sum of a harmonic form plus an exact form on $\Sigma_d$. In fact it is the harmonic piece that is important 
Later we will see later that the harmonic piece is the important part of the source.
In the case that $\Sigma_d$ is $\mathbb{R}^d$ or a torus, for example, 
we could take an independent basis of sources to be the $d$ one-forms
$E_i dx^i$ (no sum on $i$) and the $d$ one-forms
$\zeta_i dx^i$ (no sum on $i$) with constant $E_i,\zeta_i$.

The functions $M,N$ in \eqref{pertansatz} depend on $(r,x^i)$ and we will
fix them in terms of the background black hole solution via
\begin{align}\label{mandn}
M=GU\,,\qquad
N=a_t\,.
\end{align}
This is assumed, in addition to \eqref{closed}, in order to solve the time dependence of the equations of motion at linear order.
In general, this ansatz for the perturbation contains some residual gauge symmetry, 
which, for our purposes, will not need to be fixed.

At the $AdS_D$ boundary, as $r\to \infty$, we will demand that the fall-off of $\delta g_{\mu\nu}$,$\delta a_\mu$, $\delta\phi$ is 
such that the only applied sources are parametrised by $(E,\zeta)$.
The asymptotic fall-offs of $\delta g_{\mu\nu}$,$\delta a_\mu$, $\delta\phi$ are associated with currents and other expectation values that are produced by the sources. The resulting currents will be our primary interest in
this paper.

At the black hole horizon, as $r\to 0$, regularity implies that we must have
\begin{align}\label{eq:nh_exp}
\delta g_{tt}&=U\left(r \right)\,\left(\delta g^{(0)}_{tt}\left(x\right)+{\cal O}(r) \right),\quad
\delta g_{rr}=\frac{1}{U}\,\left( \delta g_{rr}^{(0)}\left(x\right)+{\cal O}(r)\right),\nn
\delta g_{ij}&=\delta g_{ij}^{(0)}\left(x\right)+{\cal O}(r),
\quad
\delta g_{tr}=\delta g_{tr}^{(0)}\left(x\right)+{\cal O}(r)\,,\nn
\delta g_{ti}&=\delta g_{ti}^{(0)}\left(x\right)-M\,\zeta_{i}\,\frac{\ln r}{4\pi T}+{\cal O}(r),\quad
\delta g_{ri}=\frac{1}{U}\,\left( \delta g_{ri}^{(0)}\left(x\right)+{\cal O}(r) \right)\,,\nn
\delta a_{t}&=\delta a_{t}^{(0)}\left(x\right)+{\cal O}(r),\quad
\delta a_{r}=\frac{1}{U}\,\left(\delta a_{r}^{(0)}\left(x\right)+{\cal O}(r)\right)\,,\nn
\delta a_{i}&=\frac{\ln{r}}{4\pi T}(-E_i+N\zeta_i)+{\cal O}(r^0)\,,
\end{align}
with the following constraints on the leading functions of $x$:
\begin{align}\label{eq:nh_constr}
&\delta g_{tt}^{(0)}+\delta g_{rr}^{(0)}-2\,\delta g_{rt}^{(0)}=0,\qquad \delta g^{(0)}_{ri}=\delta g^{(0)}_{ti},\qquad
\delta a_{r}^{(0)}=\delta a_{t}^{(0)}\,. 
\end{align}
It is worth emphasising that the logarithm terms that appear in \eqref{eq:nh_exp} are a direct consequence of 
the applied sources $(E,\zeta)$. For the scalar field we have $\delta\phi=\delta\phi^{(0)}(x)+ \mathcal{O}(r)$.

In the Kruskal coordinates, at leading order in the expansion in $r$ we have
\begin{align}
\delta(ds^2)\sim&\,\,\,4\pi T r \delta g^{(0)}_{tt}dv^2+2 dv dr(-\delta g_{tt}^{(0)}+\delta g_{tr}^{(0)})+2\delta g^{(0)}_{ti}dv dx^i+\delta g_{ij}^{(0)} dx^i dx^j\,,
\nn
&+2(vG^{(0)} \zeta_i-\delta g_{ti}^{(1)}+\delta g_{ri}^{(1)}) dr dx^i
+\frac{1}{4\pi T}\left(\delta g_{tt}^{(1)}+\delta g_{rr}^{(1)}-2\,\delta g_{rt}^{(1)}\right)dr^2\,,\nn
\delta A\sim&\,\,\,\delta a_t^{(0)} dv+\left(\delta a^{(0)}_i-v E_i\right)dx^i+\frac{1}{4\pi T}(\delta a^{(1)}_r-\delta a_t^{(1)}) dr\,.
\end{align}
Note that to obtain the leading order pieces in the perturbed field strength one should calculate the field strength first and then take the limit $r\to 0$.

\subsection{Electric current}
We define the bulk electric current density via
\begin{align}
J^i=\sqrt{-g}Z(\phi)F^{ir}\,.
\end{align}
When evaluated at the $AdS_D$ boundary we obtain $J^i|_\infty$ which is the electric current density of the dual field theory as explained in 
appendix \ref{holcurr}. 

In the background geometry we have $J^i=0$. At linearised order for the perturbed black holes we have
\begin{align}\label{jexpression}
J^i=\frac{\sqrt{g_d}g^{ij}_d}{(FG)^{1/2}}GUZ(\phi)\Bigg(\partial_ja_t \frac{\delta g_{rt}}{GU}-\partial_r a_t(\frac{\delta g_{jt}}{GU}-\frac{tM\zeta_j}{GU})
+\partial_j\delta a_r-(\partial_r\delta a_j+t \partial_rN\zeta_j)\Bigg)\,,
\end{align}
and we see that the time-dependence drops out because of \eqref{mandn}.

The gauge equations of motion given in \eqref{geneoms} can be written in the form
\begin{align}\label{jandfeoms}
\partial_iJ^i&=0\,,\nn
\partial_rJ^i&=\partial_j\left(\sqrt{-g}Z(\phi)F^{ji}\right)\,,
\end{align}
as well as 
\begin{align}\label{jandfeoms2}
\partial_r\left(\sqrt{-g}Z(\phi)F^{rt}\right)&=-\partial_i\left( \sqrt{-g}Z(\phi)F^{it}\right)\,.
\end{align}

For later use we also note that the perturbation satisfies, at linearised order,
the condition
\begin{align}\label{dxsf}
d(i_{k}*Z(\phi)F)=0\,,
\end{align}
where $k=\partial_t$.
Indeed this easily follows by writing the components of the $d-1$ form as
\begin{align}\label{starzf}
i_{k}*Z(\phi)F=&(-1)^{d-2}\Bigg[\tfrac{1}{(d-1)!}\epsilon(i_i\dots i_{d-1}j)J^jdx^{i_1}\wedge \dots \wedge dx^{i_{d-1}}\nn
&+\tfrac{1}{2(d-2)!}\epsilon(i_1\dots i_{d-2}jk) \sqrt{-g}Z(\phi)F^{jk}dx^{i_1}\wedge\dots dx^{i_{d-2}}\wedge dr
\Bigg]\,,
\end{align}
where $\epsilon$ is the alternating symbol with $\epsilon(1\dots d)=1$ 
and then using \eqref{jandfeoms}. Note that in the special case when $\zeta =0$ we have that $k=\partial_t$ is a Killing vector with ${\cal L}_{k} F={\cal L}_{k} \phi=0$. It is then very easy to establish \eqref{dxsf}. When $\zeta\ne 0$, $k$ is no longer
a Killing vector and furthermore ${\cal L}_{k} F\ne 0$. Nevertheless, at linearised order, we still have
\eqref{dxsf} as we have just shown.

\subsection{Heat current}

We now define the bulk heat current. To do so 
we want to identify equations of motion involving the metric perturbation 
that have a similar structure to the gauge equations of motion. We do this using\footnote{Heuristically, one can view this as a Kaluza-Klein reduction on the time direction. Alternatively, the analysis is inspired by derivations of the first law
of black hole mechanics e.g. \cite{Gauntlett:1998fz}.} 
the vector $k\equiv \partial_t$. The procedure is slightly
subtle when $\zeta\ne 0$ since in this case $\partial_t$ is no longer a Killing vector. We proceed as follows.
Consider a general vector $k$ which satisfies
\begin{align}\label{kcon}
\nabla_\mu k^\mu=0,\qquad
\nabla_\mu\nabla^{(\mu}k^{\nu)}=\alpha k^\nu\,,
\end{align}
for some function $\alpha$. Note, in particular that a Killing vector satisfies these conditions
with $\alpha=0$.
The conditions \eqref{kcon} imply that
\begin{align}
\nabla_\mu\left(\nabla^{[\mu}k^{\nu]}\right)=(-R^\nu{}_\sigma +\alpha\delta^\nu_\sigma)k^\sigma\,.
\end{align}
We next write $\varphi=i_{k}A$ and $i_{k}F=d\theta +\psi$, with $\psi$ a one-form and $\theta$ a globally defined function.
In the special case that ${\cal L}_k F=0$ we have $d\psi=0$.
We now define\footnote{This definition slightly differs from the definition used in \cite{Donos:2014cya,Donos:2014yya}. The expression here has the advantage that it is globally defined.} 
the two-form $G$:
\begin{align}\label{geedefone}
G^{\mu\nu}=-2\nabla^{[\mu}k^{\nu]}-\frac{2Z(\phi)}{D-2} k^{[\mu}F^{\nu]\sigma}A_\sigma-\frac{1}{(D-2)}\,\left[\left(3-D\right)\,\theta+\varphi \right]Z(\phi)\,F^{\mu\nu}\,.
\end{align}
If we assume that ${\cal L}_k\phi=0$, using the equations of motion \eqref{geneoms}, we can deduce that
\begin{align}\label{geeeqn}
\nabla_{\mu}G^{\mu\nu}=\left(\alpha+\frac{2V}{D-2} \right)\,k^{\nu}-\frac{3-D}{D-2}Z(\phi)F^{\nu\rho}\psi_{\rho}-\frac{1}{D-2}\,Z(\phi)A_{\rho}L_k(F^{\nu\rho}) \,.
\end{align}
For our setup, with $k=\partial_t$ and working at linearised order, we can choose $\theta=-\varphi$ and we have
\begin{align}
\alpha&=-\nabla^{(d)}_i(g_{(d)}^{ij}\zeta_j)-\frac{1}{2}\partial_i\log (G^3 F) g_{(d)}^{ij}\zeta_j\,,\nn
\varphi&=-\theta=a_{t}+\delta a_{t}\,,\qquad \psi=-E_{i}\,dx^{i}+a_{t}\,\zeta_{i}\,dx^{i}\,.
\end{align}

We now define the bulk heat current density via
\begin{align}
Q^i=\sqrt{-g}G^{ir}\,.
\end{align}
When evaluated at the $AdS_D$ boundary, we show in appendix \ref{holcurr} that 
$Q^i|_\infty$ is the time-independent part of the heat current density of the dual CFT:
\begin{align}\label{defq}
\bar G^{1/2}\sqrt{\bar g_d} (\bar Gt^{ti}-\mu j^{i})&= Q^i|_\infty- t \bar G^{3/2} \sqrt{\bar g_d} t^{ij} \zeta_{j} \,.
\end{align}
Here $t^{ti}$, $j^i$ are the expectation values of the holographic stress tensor and current vector, with e.g. $J^i|_\infty=\bar G^{1/2}\sqrt{\bar g_d} j^i$ 
The precise
combination that appears on the left hand side in \eqref{defq}
is the operator that is sourced by $-t\zeta_i$ and is, by definition, what we call the heat
current density. Notice that in the case when the holographic lattice has no spatially dependent sources for the
metric this reduces to the standard expression $t^{ti}-\mu j^{i}$.
The time dependent piece on the right hand side is associated with the static susceptibility for the
heat current two point function (see appendix C of \cite{Donos:2014cya}).

In the background geometry we have $Q^i=0$. At linearised order for the perturbed black holes we have
\begin{align}\label{defqueue}
Q^i=\frac{G^{3/2}U^2}{F^{1/2}}\sqrt{g_d}g^{ij}_d\left(\partial_r\left(\frac{\delta g_{jt}}{GU}\right)-\partial_j\left(\frac{\delta g_{rt}}{GU}\right)
\right)-a_t J^i\,.
\end{align}
\begin{align}\label{Qeqns}
\partial_iQ^i&=0\,,\nn
\partial_rQ^i&=\partial_j\left(\sqrt{-g}G^{ji}\right)\,,
\end{align}
as well as
\begin{align}\label{secqeqn}
\partial_r\left(\sqrt{-g}G^{rt}\right)+\partial_i\left( \sqrt{-g}G^{it}\right)
=\sqrt{-g}\left(
\left(\alpha+\frac{2V}{D-2} \right)-\frac{3-D}{D-2}Z(\phi)F^{t j}\psi_{j}
\right)\,.
\end{align}
Note that \eqref{secqeqn} includes an equation for the background as well as the linearised
perturbation.
We also record here that
\begin{align}
\sqrt{-g}G^{ij}=-(GF)^{1/2} \sqrt{g_d}g^{ik}_d g^{jl}_d
\Bigg((UG)\partial_k\left(\frac{\delta g_{lt}}{GU}\right)+&Z(\phi)a_t\left(
\partial_k a_t\left(\frac{\delta g_{lt}}{GU}\right)+\partial_k\delta a_l\right)
&-k\leftrightarrow l
\Bigg)\,,
\end{align}
and we note that the time dependence drops out because of the conditions 
\eqref{closed} and \eqref{mandn}.

Finally, following the discussion for the electric currents, with $k=\partial_t$ we conclude that
\eqref{Qeqns} implies 
\begin{align}\label{dxsf2}
d(i_{k}*G)=0\,.
\end{align}

\subsection{Currents at the horizon}\label{curhor}
We now obtain expressions for the electric and the heat current densities by expanding at the black hole horizon.
We find:
\begin{align}\label{eq:J_hor}
J^i_{(0)}&\equiv\left.J^i\right|_H =Z(\phi^{(0)})\sqrt{g_{(0)}}g^{ij}_{(0)}\left(\left(\partial_j\delta a_t^{(0)}+E_j\right)-a^{(0)}_t\delta g^{(0)}_{jt}\right)\,,\nn
Q^i_{(0)}&\equiv\left.Q^i\right|_H =-4\pi T\sqrt{g_{(0)}}g^{ij}_{(0)}\delta g^{(0)}_{jt}\,.
\end{align}
From the first equations in \eqref{jandfeoms} and \eqref{Qeqns} we immediately obtain
\begin{align}\label{divjq}
\partial_iJ^i_{(0)}=0\,,\qquad
\partial_iQ^i_{(0)}=0\,,
\end{align}
which give two equations for a subset of the perturbations at the horizon. We can obtain a closed
system of equations, which are the generalised Stokes equations, by considering
the second equation of \eqref{Qeqns}. We explain how this can be achieved in appendix \ref{otherder}.
The same system of equations can also be obtained, in a more illuminating manner,
by evaluating the Hamiltonian, momentum and Gauss law constraints on the black hole horizon, as we now discuss.

\subsection{Constraints at the horizon}\label{consdesc}
We carry out a Hamiltonian decomposition of the equations of motion using
a radial decomposition in appendix \ref{radham}. The momentum constraints and the Gauss-law constraints,
can be written in the form $H^{\nu}=C=0$ where
\begin{align}
H^{\nu}=&-2\sqrt{-h}\,D_{\mu}\left((-h)^{-1/2}\pi^{\mu\nu}\right)+h^{\nu\sigma}f_{\sigma\rho} \pi^\rho
-h^{\nu\sigma}a_\sigma \partial_\rho  \pi^\rho
+h^{\nu\sigma}\partial_\sigma\phi\pi_\phi \,,\nn
C=&\partial_\mu\pi^{\mu}\,,
\end{align}
with
\begin{align}
\pi^{\mu\nu}&=-\sqrt{-h}\,\left( K^{\mu\nu}-K\,h^{\mu\nu}\right)\,,\notag\\
\pi^\mu
& =\sqrt{-h} Z F^{\mu\rho}n_\rho\,,\nn
\pi_\phi&=-\sqrt{-h}n^\nu\partial_\nu\phi\,.
\end{align}
Here $n$ is the unit norm normal vector, $h_{\mu\nu}=g_{\mu\nu}-n_\mu n_\nu$ is the induced metric, 
$K_{\mu\nu}=\frac{1}{2}{\mathcal L}_n h_{\mu\nu}$ is the extrinsic curvature
and $K=g^{\mu\nu}K_{\mu\nu}$. In addition $D_\mu$ is the Levi-Civita connection with respect to $h_{\mu\nu}$,
$b_\mu=h_\mu{}^\nu A_\nu$ and $f_{\mu\nu}=\partial_\mu b_\nu-\partial_\nu b_\mu$.

We want to analyse these constraints for the perturbed metric on a surface of constant $r$, near the horizon, and then take the limit
$r\to 0$. In local coordinates we have $n=N dr$ where $N$ is the lapse function. We immediately notice that
\begin{align}
\pi^\mu
 =\sqrt{-h} Z F^{\mu r}n_r=J^\mu\,,
\end{align}
and hence the Gauss-Law constraint $C=0$ is simply $\partial_\mu J^\mu=0$. Evaluated
at the horizon we obtain $\partial_iJ^i_{(0)}=0$ as in \eqref{divjq}. Turning to the momentum constraint, which we discuss further in appendix \ref{navst}, 
we find that evaluating $H_t=0$ as an expansion at the horizon gives $\partial_i Q^i_{(0)}=0$ as in \eqref{divjq}. Next, $H_i=0$ evaluated
at the horizon gives the extra equation mentioned above which, combined with 
$\partial_iJ^i_{(0)}=\partial_i Q^i_{(0)}=0$ gives the Stokes system of equations which we summarise in the next subsection. Finally, we note that the leading order term
of the Hamiltonian constraint, which is explicitly given in \eqref{eq:H_constraint}, 
also gives the condition $\partial_i Q^i_{(0)}=0$.

\subsection{Generalised Stokes equations at the horizon}
We can now summarise the closed system of equations that we have shown a subset
of the linearised perturbations must satisfy at the black hole horizon. 
The black hole horizon is as in \eqref{eq:DC_ansatz}, \eqref{nhexpbh} and \eqref{gandf}. The perturbation at 
the horizon is given as in \eqref{eq:nh_exp},\eqref{eq:nh_constr} and it is illuminating to now introduce the following
notation:
\begin{align}\label{deffquants}
v_{i}\equiv -\delta g_{it}^{(0)},\qquad w\equiv \delta {a}_{t}^{(0)},\qquad p\equiv -4 \pi T\frac{\delta g_{rt}^{(0)}}{G^{(0)}}-\delta g_{it}^{(0)}g^{ij}_{(0)}\nabla_{j}\,\ln G^{(0)}\,.
\end{align}
These $d+2$ unknowns satisfy the following $d+2$ linear system of partial differential equations:
\begin{align}
\nabla_{i} v^{i}&=0\,, \label{eq:v_eq2}\\
\nabla_i(Z^{(0)}\nabla^i w)+v^{i}\,\nabla_{i}\left({Z^{(0)}a_{t}^{(0)}}\right)&=-\nabla_{i}(Z^{(0)} E^{i})\,,\label{eq:w_eq2}\\
-2\,\nabla^{i}\nabla_{\left( i \right. }v_{\left. j\right)}-{Z^{(0)}a_{t}^{(0)}}\nabla_j w
+\nabla_j\phi^{(0)}\nabla_i\phi^{(0)}v^{i}
+\nabla_{j}\,p&=4\pi T\,\zeta_{j}
+{Z^{(0)}a_{t}^{(0)}} E_j\,,\label{eq:V_neutral2}
\end{align}
where the covariant derivatives $\nabla$ in this subsection are with respect to the metric, $g^{(0)}_{ij}$, on the black hole horizon $\Sigma_d$, and all indices are being raised and lowered
with this metric. The first two equations are simply $\partial_i Q^i_{(0)}=\partial_i J^i_{(0)}=0$, where $Q^i_{(0)}, J^i_{(0)}$ are the heat current and electric current densities at the horizon, respectively:
\begin{align}\label{eq:J_hor2}
Q^i_{(0)} &=4\pi T\sqrt{g_{(0)}}v^j\,,\nn
J^i_{(0)}&=\sqrt{g_{(0)}}g^{ij}_{(0)}Z^{(0)}\left(\partial_j w+{a^{(0)}_t}v_j+E_j\right)\,.
\end{align}
It is helpful to note that in the third equation we can also write
\begin{align}\label{rvisc}
2\,\nabla^{i}\nabla_{\left( i \right. }v_{\left. j\right)}=\nabla^2v_j+R_{ji} v^i\,.
\end{align}

We emphasise that by evaluating the constraints at the horizon we obtain a system of 
equations for a {\it subset} of the linear perturbation, namely, $\delta g_{it}^{(0)}, \delta {a}_{t}^{(0)}, \delta g_{rt}^{(0)}$,
and we obtain a closed system for this set. Furthermore, the equations we have obtained are a generalisation of the forced Stokes equations for a charged fluid on the curved black hole horizon. 
Indeed in the special case of electrically neutral black hole horizons with $a^{(0)}_t=w=E=0$ and in addition constant $\phi^{(0)}$, 
the equations are simply the Stokes equations with fluid velocity $v_i$, pressure $p$ and forcing
term given by the closed one-form $4\pi T\zeta$. The curvature of the horizon gives rise to an extra viscosity term as in \eqref{rvisc}.
In the general case we have a charged fluid with scalar potential $w$ and an additional forcing term 
given by the closed one-form $E$. It is also interesting to note that the scalar field is giving viscosity terms of the form $\nabla^j\phi^{(0)}\nabla^i\phi^{(0)}v_{i}$.
We will see in section 4 how these extra terms play a direct role in determining the DC conductivity. 
We emphasise that we have not taken any hydrodynamical limit in obtaining these equations.

We now establish a number of interesting properties of this set of equations.
Firstly, by taking the divergence of \eqref{eq:V_neutral2} and using \eqref{eq:v_eq2}, \eqref{eq:w_eq2} we obtain the ``pressure Poisson equation"
\begin{align}\label{poisseq}
\nabla^2p=
\nabla_j\Bigg(2R^j{}_k v^k+{Z^{(0)}a^{(0)}}(\nabla^j w+E^j)+4\pi T \zeta^j
-\left(\nabla^j\phi^{(0)}\nabla^k\phi^{(0)}\right)v_k\Bigg)\,.
\end{align}
For a compact horizon and given background data, the pressure term is uniquely specified by $v_j, w,E_j, \zeta_j$.

Second, we multiply \eqref{eq:V_neutral2} by $v_{j}$ from the left, and then integrate over the horizon,  
and use \eqref{eq:v_eq2},\eqref{eq:w_eq2} to obtain
\begin{align}\label{positiveexp}
&\int d^{d}x\, \sqrt{g_{{0}}} \left[2\nabla^{\left( i \right. }v^{\left. j\right)} \nabla_{\left( i \right. }v_{\left. j\right)}+Z^{(0)}\left(\nabla w+E \right)^{2}
+v_i\left(\nabla^i\phi^{(0)}\nabla^j\phi^{(0)}\right)v_j\right]\nn
&\qquad\qquad=
\int d^{d}x\left( Q^i_{(0)}\zeta_i+ J^i_{(0)} E_i\right)
\end{align}
In the case of non-compact horizons, we have assumed that possible boundary terms vanish.
Observe that the left hand side is a manifestly positive quantity and this is
related to the positivity of the thermoelectric conductivities, which we discuss later. 

Third, we consider the issue of uniqueness of the equations \eqref{eq:v_eq2}-\eqref{eq:V_neutral2}. 
If we have two solutions then the difference of the functions, which we again write as $(v_i,w,p)$, will
satisfy the same equations but with vanishing forcing terms, $\zeta_{j}=E_j=0$. From \eqref{positiveexp} we immediately conclude
that
\begin{align}\label{nabivj}
\nabla^{\left( i \right. }v^{\left. j\right)} =0,\quad
\nabla^iw=0,\quad v^i\nabla_i\phi^{(0)}=0\,.
\end{align}
We also have $\mathcal{L}_{v}a_{t}^{(0)}=0$ from \eqref{eq:w_eq2} and $\nabla_i p=0$ from \eqref{eq:V_neutral2}.
We conclude that the solution space of equation \eqref{eq:V_neutral2} is unique up to Killing vectors of the horizon metric, with $p,w$ constant. We then have $\delta g_{rt}=(4\pi T)^{-1}\,\mathcal{L}_{v}G^{(0)}$ plus a constant. 
This result agrees with the intuition that one should be able to boost along the orbits of Killing vectors to obtain a solution with momentum at the horizon.

Fourth, we observe that when $(E,\zeta)$ are exact forms, $(E,\zeta)=(de,dz)$ with $e,z$ globally defined
functions on $\Sigma_d$, we can solve the equations \eqref{eq:v_eq2}-\eqref{eq:V_neutral2} by taking
$w=-e$ and $p=4\pi Tz$, plus possible constants, and $v^i=0$. We observe that this solution gives no
contribution to the current densities \eqref{eq:J_hor2} at the horizon. We will see later that this solution gives no
contribution to suitably averaged currents at the $AdS$ boundary and hence no contribution to the DC thermoelectric conductivity {\it i.e.} the DC conductivity is determined by the harmonic part of $E$ and $\zeta$. A basis for the non-trivial
part of these sources is thus given by a basis for the first cohomology group 
of $\Sigma_d$.

Fifth, we point out that the fluid equations that we have obtained can
be obtained by varying the following functional:
\begin{align}\label{ellfluid}
L=\int d^{d}x\, &\sqrt{g_{{0}}} \Bigg[-\nabla^{\left( i \right. }v^{\left. j\right)} \nabla_{\left( i \right. }v_{\left. j\right)}
-\frac{1}{2}(v^i\nabla_i\phi)^2
+p(\nabla_i v^i)
+\frac{1}{2}Z^{(0)}\left({a^{(0)}}v+\nabla w\right)^2\nn
&-\frac{1}{2}Z^{(0)}a^{(0)2}v^2+4\pi T\zeta_i v^i+Z^{(0)}E_i\left( {a^{(0)}}v^i+\nabla^i w \right)+\frac{1}{2}Z^{(0)}
E_iE^i\Bigg]\,,
\end{align}
and we remind the reader that the covariant derivative $\nabla$
in this subsection is with respect to the metric $g^{{0}}_{ij}$.
Varying with respect to the pressure, which is a Lagrange multiplier, gives the incompressibility condition.
Varying with respect $v$ and $w$ then gives the remaining Stokes equations. It is also interesting to note that
if we vary with respect to $E_i$ and $\zeta_i$ then we get the currents at the horizon
$J^i_{(0)}$ and $Q^i_{(0)}$, respectively. On shell we therefore can deduce, for example, that
\begin{align}
\frac{\delta J^i_{(0)}}{\delta\zeta_j}
=\frac{\delta Q^j_{(0)}}{\delta E_j}\,.
\end{align}
This is a kind of Onsager reciprocal relation for the currents at the horizon. After considering the current fluxes, to be 
described in the next subsection, we obtain Onsager relations for the DC conductivitities.

Finally, we comment on the fact that, {\it locally}, the sources can be eliminated from the 
Stokes equations \eqref{eq:v_eq2}-\eqref{eq:V_neutral2}. Indeed since
the sources $E,\zeta$ are closed, locally we can write $E=de$, $\zeta=dz$ and after defining
$\tilde w=w+e$, $\tilde p=p-4\pi T z$ we have
\begin{align}
\nabla_{i} v^{i}&=0\,, \\
\nabla_i(Z^{(0)}\nabla^i \tilde w)+v^{i}\,\nabla_{i}\left({Z^{(0)}a_{t}^{(0)}}\right)&=0\,,\\
-2\,\nabla^{i}\nabla_{\left( i \right. }v_{\left. j\right)}-{Z^{(0)}a_{t}^{(0)}}\nabla_j \tilde w
+\nabla_j\phi^{(0)}\nabla_i\phi^{(0)}v^{i}
+\nabla_{j}\,\tilde p&=0
\,.
\end{align}
It is important to emphasise that now $\tilde w$ and $\tilde p$ are not globally defined functions
on the black hole horizon. For example, if the horizon was a torus with $x^i=x^i+L_i$,
and the source $E= c dx^1$ for some constant $c$, then $\tilde w$ would satisfy the twisted
boundary condition $\tilde w(x^1+L_1)=\tilde w+cL_1$. Note that this would give extra contributions to
\eqref{positiveexp}. It is therefore most natural to work with the formulation with sources, and we will do so
in the sequel.

It is worth noting, however, that the sources can also be removed, locally, from the full linearised perturbation.
Indeed, suppose we carry out the gauge transformation
$A\to A= d (te)+ B$ and in addition change the time coordinate via
$t=\tilde t(1-z)+C$, where $B,C$ are functions independent of the time coordainte. 
We then find that at linearised order we obtain the same perturbed
ansatz with vanishing sources and 
\begin{align}
&\delta g_{\tilde t\tilde t}=\delta g_{tt}+2 UGz,\quad
\delta g_{\tilde t r}=\delta g_{tt}- UG\partial_r C\quad
\delta g_{\tilde t i}=\delta g_{tt}-UG\partial_i C\nn
\qquad
&\delta a_{\tilde t}=\delta a_t +e -a_t z,\quad \delta a_r=\delta a_r+\partial_r B,
\quad \delta a_i=\delta a_i+\partial_i B
\end{align}
We choose $B,C$ to vanish suitably fast at the AdS boundary and at the horizon we choose $B=\ln r/(4\pi T) e+\dots$ and $C =-\ln r/(4\pi T) z+\dots$.
Evaluating at the horizon we see that in the new coordinates and gauge
we have induced $w\to \tilde w$, $p\to \tilde p$
and $\delta g^{(0)}_{tt}\to \delta g^{(0)}_{tt}+2 G^{(0)}z$. 

\subsection{The DC thermoelectric conductivity}
For a given set of sources $(E,\zeta)$ we can solve the 
Stokes equations \eqref{eq:v_eq2}-\eqref{eq:V_neutral2} at the black hole horizon and hence
obtain expressions for the electric and heat current densities
at the black hole horizon. From this data we would
like to deduce something about the current densities at
the holographic boundary as a function of $(E,\zeta)$.
The radial dependence of the current densities are given by
\eqref{jandfeoms} and \eqref{Qeqns}. For some simple special
classes of black hole solutions the current densities $J^i,Q^i$ are independent of the radial
coordinate. This occurs for the Q-lattices and the holographic lattices that depend on just one of the spatial dimensions,
for example. However, for general classes of black holes $J^i,Q^i$ will depend on the radial direction.

On the other hand, remarkably, we can always define ``current flux densities" $\bar J^i,\bar Q^i$ which are independent of $r$ and hence we can obtain the associated DC conductivity.
To consider a concrete example, we assume that we are in $D=4$ with a periodic holographic lattice and $\Sigma_d=\mathbb{R}^2$ or a two-torus.
In particular, the lattice deformations $\bar G(x), \bar g_{ij}(x)$, $\mu(x)$ and $\bar\phi(x)$
in \eqref{asymptsol} are all periodic functions of the spatial coordinates $x^i$ with
period $L_i$.
We define the following current flux densities
\begin{align}\label{avecur}
\bar J^1\equiv \frac{1}{ L_2}\int J^1 dx^2\,,\qquad
\bar J^2\equiv \frac{1}{ L_1}\int J^2 dx^1\,,
\end{align}
where $\bar J^1$ and $\bar J^2$ is the current flux density through the $x_2$ and $x_1$ planes, respectively.
We define $\bar Q^i$ in a similar way. 
We can then immediately deduce from \eqref{jandfeoms}, \eqref{Qeqns}
that $\partial_r\bar J^i=\partial_r\bar Q^i=0$, which is simply Stokes theorem in the bulk.
Notice that $\bar J^i,\bar Q^i$ are also independent of the $x^i$ coordinates and hence they are just constants.
Similarly in $D=5$ with a periodic lattice on $\mathbb{R}^{1,3}$
we would define the following constant current flux densities
\begin{align}\label{avecur2}
\bar J^1\equiv \frac{1}{ L_2L_3}\int  dx^2dx^3 J^1\,,\quad
\bar J^2\equiv -\frac{1}{ L_3L_1}\int  dx^1dx^3 J^2\,,\quad
\bar J^3\equiv \frac{1}{ L_1L_2}\int  dx^1dx^2 J^3\,,
\end{align}

The current flux densities of the boundary theory are, by definition, $\bar J^i,\bar Q^i$ evaluated at the
AdS boundary. In order to evaluate the DC conductivity matrix we want to relate these to constant sources $\bar E_i,\bar\zeta_i$ at the
AdS boundary via
\begin{align}\label{bigform2}
\left(
\begin{array}{c}
\bar J^i\\\bar Q^i
\end{array}
\right)=
\left(\begin{array}{cc}
\sigma^{ij} & T\alpha^{ij} \\
T\bar\alpha^{ij} &T\bar\kappa^{ij}   \\
\end{array}\right)
\left(
\begin{array}{c}
\bar E_j\\ \bar\zeta_j
\end{array}
\right)\,.
\end{align}
We have just shown that $\bar J^i,\bar Q^i$ are constant and so their value at the AdS boundary is the same as at the black hole
horizon, and that in turn these are fixed by the closed forms $E,\zeta$ evaluated at the horizon by solving the Stokes equations \eqref{eq:v_eq2}-\eqref{eq:V_neutral2}. Continuing with the case that $\Sigma_d=\mathbb{R}^d$ or a torus, we 
can take an independent basis of sources to be the $d$ one-forms
$\bar E_i dx^i$ (no sum on $i$) and the $d$ one-forms $\bar \zeta_i dx^i$ (no sum on $i$) with constant $\bar E_i,\bar\zeta_i$ and this defines the DC conductivity matrix \eqref{bigform2}. 

Note that an equivalent way to characterise the constant sources $\bar E_i,\bar\zeta_i$ at the AdS boundary is to write a general closed form source on $\Sigma_d=\mathbb{R}^d$ or a torus, as $E=\bar E_i dx^i + de$, where $e$ is a periodic function and then extract $\bar E_i$ by
integrating $E$ over an appropriate basis of one-cycles. It is worth emphasising that in the paragraph following \eqref{nabivj} we noted that it is only the harmonic part of the sources that contribute to the currents at the horizon and hence this
procedure gives the same DC conductivity.

After substituting \eqref{bigform2} in \eqref{positiveexp},
we can now explicitly
see that the positivity of the left hand side of equation \eqref{positiveexp} implies that the thermoelectric DC conductivity is a positive definite matrix. 
Continuing on from the discussion following \eqref{ellfluid} we can deduce that the thermoelectric matrix
is symmetric.

We will discuss the DC conductivity matrix when $\Sigma_d$ is
not $\mathbb{R}^d$ or a $d$-dimensional torus in the next subsection.

We have presented a procedure for calculating the DC conductivity of
the boundary theory in terms of a calculation at the black hole horizon. One might wonder if the calculation
could also be done at any constant radial hypersurface. In fact this cannot be done since evaluating the constraint
equations on a constant $r$ hypersurface with $r\not\to 0$ will not lead to a closed system of equations for
a subset of the linear perturbation and hence we cannot obtain the current fluxes.

\subsection{Perspective using forms}\label{gens}
We have focussed on a particular class of black holes given in \eqref{eq:DC_ansatz}, with a single black hole horizon, and with perturbation given
in \eqref{pertansatz} with \eqref{closed}, \eqref{mandn}. We have also
focussed on black holes for which $\Sigma_d$ is topologically either $\mathbb{R}^d$ or a $d$-dimensional torus.
In this section we briefly discuss the DC conductivity calculation using the language of forms, which illuminates some global issues
as well as revealing generalisations for $\Sigma_d$ with other topologies.

One key point 
is that the two-forms $F$ and $G$ for the perturbed metric satisfy, at linearised order, the following closure conditions (see \eqref{dxsf}, \eqref{dxsf2}):
\begin{align}\label{closure}
d(i_{k}*Z(\phi)F)=0\,,
\qquad
d(i_{k}*G)=0\,,
\end{align}
where $k=\partial_t$.
These conditions are valid without $k$ being a Killing vector, but instead satisfying the
weaker conditions in \eqref{kcon}. 

We now observe that for any two-forms satisfying \eqref{closure} we can define a natural set of 
current fluxes. Specifically, at the deformed $AdS_D$ boundary and at fixed $t$, we let $C_a$,
$a=1,\dots, b_{d-1}(\Sigma_d)$, 
be a basis of $d-1$ closed cycles on $\Sigma_d$, where $b_{d-1}(\Sigma_d)$ is the $(d-1)$ Betti number of $\Sigma_d$. 
We can then define the current fluxes through these cycles via
\begin{align}\label{defjay}
\bar J^a\equiv - \int_{C_a}i_{k}*Z(\phi)F\,,
\qquad
\bar Q^a\equiv - \int_{C_a}i_{k}*G\,.
\end{align}
We now consider a $d$-dimensional surface $S$ 
in the bulk spacetime which has boundary $C_a$ at the $AdS$ boundary and possibly another boundary at a black hole
horizon. Then since the integrand in \eqref{defjay} is closed, we deduce that these current fluxes are also equal to their values at  the black hole horizon.  If the cycle $C_a$ is contractible in the bulk, then the current flux would necessarily have to be zero.

For the special cases of periodic lattices in $D=4$ and $5$ spacetime dimensions, for which $\Sigma_d$ is topologically $\mathbb{R}^2$ and $\mathbb{R}^3$, respectively,  using \eqref{starzf} we immediately see that the definition \eqref{defjay} agrees 
with the definitions\footnote{Note that in \eqref{avecur}, \eqref{avecur2} we have divided by suitable $L_i$ in order to obtain current flux {\it densities}.} 
given in 
\eqref{avecur}, \eqref{avecur2} after choosing an obvious basis
of one and two cycles, respectively. For these cases, the number of current fluxes is the same as the dimension of $\Sigma_d$. However, this is not the case for more general $\Sigma_d$. In $D=4$, for example, we can envisage black holes in which $\Sigma_2$ is a Riemann surface with genus $g>1$, and it is possible
to define $2g$ current fluxes. There are 
many more possibilities for solutions in
$D=5$. We also note that when $\Sigma_d$ is a sphere,
which is relevant for solutions associated with deformations of global $AdS$, 
these current fluxes are all trivial since $b_{d-1}(S^d)=0$.

The above comments were based on general two-forms satisfying \eqref{closure}. A second key point in our derivation of the DC conductivity is that
the two-forms were constructed with specific source terms parametrised by the one-forms $E,\zeta$. For the class of solutions that
we considered we assumed there was a single black hole horizon with the same topology $\Sigma_d$ as the spatial boundary of the 
deformed $AdS$ space. 
In order to satisfy \eqref{closure} it was necessary to take the one forms 
$E,\zeta$ to be closed one-forms on $\Sigma_d$ and independent of the radial coordinate. 
Corresponding to the basis of $(d-1)$-cycles, $C_a$, we can define a basis of harmonic one-forms, $\phi^a$, on $\Sigma_d$ by
Poincar\'e duality. We can then write $E=\bar E_a\phi^a +de$ and $\zeta =\bar\zeta_a\phi^a +dz$ with
constant $\bar E_a,\bar\zeta_a$. Recalling the discussion in the paragraph following \eqref{nabivj},  
in solving the Stokes equations at the horizon in order to obtain the currents at the horizon only the harmonic
part of the sources, $\bar E_a\phi^a$, $\bar\zeta_a\phi^a$ are important.  We therefore can
relate the $b_{d-1}(\Sigma_d)$ independent constant source terms
to the $b_{d-1}(\Sigma_d)$ current fluxes at the horizon,
after solving the Stokes equations, and hence to the 
$b_{d-1}(\Sigma_d)$ current fluxes at the $AdS_D$ boundary. 
This procedure give rise to thermoelectric 
conductivity matrices $\sigma^{ab}$, $\alpha^{ab}$, $\bar\alpha^{ab}$ and
$\bar\kappa^{ab}$, all of which are $b_1(\Sigma_d)\times b_1(\Sigma_d)$
matrices, where we used the fact that $b_{d-1}(\Sigma_d)=b_1(\Sigma_d)$.

One can ask if this aspect of the formalism can be adopted to more general classes of black holes in which there are multiple black hole horizons
(an example of such a solution, but without spatially dependent sources, is given in \cite{Horowitz:2014gva}). This would require identifying
suitable source terms $E,\zeta$ that depend on the radial direction while still maintaining the condition \eqref{closure}. We return to this point in section \ref{disc}.

Finally, recalling \eqref{positiveexp} we note that we can write
\begin{align}
\int d^{d}x\left( Q^i_{(0)}\zeta_i+ J^i_{(0)} E_i\right)&=-\int_{\Sigma_d}  (i_{k}*Z(\phi)F)\wedge E+(i_{k}*G)\wedge \zeta\,,\nn
&=-\int_{C^{(E)}}(i_{k}*Z(\phi)F)-\int_{C^{(\zeta)}} (i_{k}*G)\,,
\end{align}
where in the first line we are integrating over any surface at constant $r$ and $t$. In the
second line
$C^{(E)}$ and $C^{(\zeta)}$ are $(d-1)$ cycles, unique up to homology, that are Poincar\'e dual 
to the closed one-forms $E$ and $\zeta$. By definition the right hand side is thus the sum of the
current fluxes $\bar J^{(E)}+\bar Q^{(\zeta)}$, through the cycles $C^{(E)}$ and $C^{(\zeta)}$, repsectively. 
The positivity of the left hand-side, which we obtain from 
\eqref{positiveexp}, is associated with the positivity of the thermoelectric DC conductivity.

\section{Examples}
In this section we examine some special examples of holographic lattices for which we can
solve the fluid equations on the horizon and hence obtain expressions for the 
DC conductivity in terms of the behaviour of the black hole solutions at
the horizon. We first discuss how extra scalar fields manifest themselves in extra terms 
in the fluid equations at the horizon and then use this to study a general class of Q-lattices.
We next analyse general holographic lattices that depend on just one spatial dimension.
Finally we examine holographic lattices that can be obtained as a perturbative expansion
about the AdS-RN black brane.

\subsection{Extra scalars and Q-lattices}
For simplicity we derived the Stokes equations for the model
given in \eqref{eq:V_neutral2} which involved a single scalar field. However, 
the generalisation to extra
scalar fields, which can parametrise a non-trivial target space manifold,
is straightforward. Specifically, if we replace
\eqref{eq:bulk_action} with several scalars, $\phi^I$, 
with the functions $V,Z$ depending on all of the scalars and
the kinetic-energy terms generalised via
\begin{align}
-\frac{1}{2}\partial\phi^2\to -\frac{1}{2}{\cal G}_{IJ}(\phi)\partial\phi^I\partial\phi^J\,,
\end{align}
then this leads to the Stokes equations as before, with the only
change in \eqref{eq:V_neutral2} given by
\begin{align}
-\nabla_j\phi^{(0)}\nabla_i\phi^{(0)}v^{i}
\to
-{\cal G}_{IJ}(\phi^{(0)})\nabla_j\phi^{I(0)}\nabla_i\phi^{J(0)}v^{i}\,.
\end{align}

With this result in hand we can now obtain previous results for the DC conductivities for Q-lattices \cite{Donos:2014uba,Donos:2014cya}. 
They key feature of the Q-lattice is that it exploits a global symmetry in the bulk to construct the black hole solutions. In the present context
we assume that the model admits $n$ global shift symmetries of the scalars:
\begin{align}
\phi^{I_{\alpha}}\rightarrow \phi^{I_{\alpha}}+\epsilon^{I_{\alpha}}\,,
\end{align}
with $\alpha=1,\dots,n$.
For example, if we had a single complex scalar field with a global $U(1)$
symmetry, then the associated shift symmetry of this type is obtained 
by parametrising the scalar manifold locally with the modulus and phase of 
the complex scalar field. This gives rise to a periodic lattice. Another example,
is a massless `axion' field with only derivative couplings.
Note that the function $Z$ must be independent of these $\phi^{I_{\alpha}}$.

The spatial coordinates $x^i$ are taken to parametrise $\mathbb{R}^d$ or possibly
a torus. The black hole solutions are then constructed based on an ansatz in which 
the scalars associated with these shift symmetries take the form
\begin{align}
\phi^{I_{\alpha}}=\mathcal{C}^{I_{\alpha}}{}_{j}\,x^{j}\,,
\end{align}
everywhere in bulk with $\mathcal{C}$ a constant $n$ by $d$ matrix. For simplicity 
of presentation we assume that all spatial coordinates are involved and hence the DC conductivity in all spatial
directions is finite. The metric, the gauge-field and the remaining scalar fields will depend on the radial direction but will be independent of the spatial coordinates $x^i$. 
The metric on the black hole horizon is flat and in addition, $Z^{(0)}$, $G^{(0)}$ and 
$a^{(0)}_t$ are all constant.

After these remarks, the fluid equations \eqref{eq:v_eq2}-\eqref{eq:V_neutral2} are solved with $v^{i}$, $p$ and $w$ all constant on the horizon. The fluid velocity is given by
\begin{align}
v^{i}&=4\pi T\,\left(\mathcal{D}^{-1} \right)^{ij}\,\left(\zeta_{j}+\frac{\rho}{Ts}\,E_{j} \right)\,,
\end{align}
with constant $E_i,\zeta_i$ and we have defined the $d\times d$ matrix:
\begin{align}
\mathcal{D}_{ij}&=G_{I_{\alpha_{1}} I_{\alpha_{2}}}\,\mathcal{C}^{I_{\alpha_{1}}}{}_{i}\,\mathcal{C}^{I_{\alpha_{2}}}{}_{j}\,.
\end{align}
Furthermore, the averaged charge density, $\rho$, defined in \eqref{rhotot},
and the entropy density, $s$, are given by 
\begin{align}
\rho=\rho_H&=\sqrt{g_{(0)}}{Z^{(0)}a_{t}^{(0)}},\qquad s=4\pi \sqrt{g_{(0)}}\,.
\end{align}
The current densities $J^i,Q^i$ are independent of the radius and are given by their
horizon values:
\begin{align}
J^{i}=& \left(\frac{s\,Z^{(0)}}{4\pi}\,g_{(0)}^{ij}+ \frac{4\pi\rho^{2}}{s}\,\left( \mathcal{D}^{-1}\right)^{ij} \right)\,E_{j}+4\pi T\rho \left( \mathcal{D}^{-1}\right)^{ij}\zeta_{j}\,,\nn
Q^{i}=&4\pi Ts \left( \mathcal{D}^{-1}\right)^{ij}\,\left(\zeta_{j}+\frac{\rho}{Ts}\,E_{j}\right)\,.
\end{align}
The DC conductivities are thus given by
\begin{align}
\sigma^{ij}=&\frac{s\,Z^{(0)}}{4\pi}g_{(0)}^{ij}+ \frac{4\pi\rho^{2}}{s}\,\left( \mathcal{D}^{-1}\right)^{ij}\,,\nn
\alpha^{ij}=&\bar{\alpha}^{ij}=4\pi\rho \,\left( \mathcal{D}^{-1}\right)^{ij}\,,\nn
\bar{\kappa}^{ij}=&4\pi T s\,\left( \mathcal{D}^{-1}\right)^{ij}\,.
\end{align}
Note that the conductivity when $Q=0$, $\sigma_{Q=0}\equiv  \sigma-T\alpha\bar\kappa^{-1}\bar\alpha$, is given by
\begin{align}
\sigma^{ij}_{Q=0}=\frac{s\,Z^{(0)}}{4\pi}g_{(0)}^{ij}\,.
\end{align}

A final point worth emphasising is that the origin of the matrix $\mathcal{D}$ appearing in the final expressions
arises from the extra terms involving the scalars in the Stokes equations, underscoring the significance of the latter.

\subsection{One-dimensional lattices}\label{sec:1d_lattices}

We now consider a class of black hole solutions with metrics on the
horizon that break translations in just one of the spatial directions.
As special sub-cases we will recover the results for
the inhomogeneous lattices with varying chemical potential studied in
\cite{Donos:2014yya} as well as the helical lattices studied in \cite{Donos:2014gya}. Recently
formulae for the DC conductivity for a scalar lattice were obtained in 
\cite{Rangamani:2015hka}
in terms the behaviour of the solution at the black hole horizon as well
as sub-leading terms. We improve upon those results by providing a new formula
that depends just on the solution at the horizon.

We assume that the horizon geometry depends on the spatial coordinate
$x$ and is independent of the remaining $d-1$ spatial coordinates which, for definiteness and without loss of generality,
we take to parametrise a torus. The moduli of this torus can depend on $x$. For simplicity we restrict our considerations to metrics on the black hole horizon of the form
\begin{align}\label{onedform}
ds^{2}_{d}=g^{(0)}_{ij}\,dx^{i}dx^{j}=\gamma(x) \,dx^{2}+ds^{2}_{d-1}\left(x\right)\,,
\end{align}
where $ds^{2}_{d-1}=g_{ab}dx^a dx^b$ is a flat metric on the torus.
We now solve the relevant system of equations \eqref{eq:v_eq2}-\eqref{eq:V_neutral2}.
The incompressibility condition \eqref{eq:v_eq2} is solved by taking
the non-vanishing components of $v^i$ to be
\begin{align}
v^x=(\gamma\,g_{d-1})^{-1/2}\,v_{0}\,,
\end{align}
with $g_{d-1}$ the determinant of the $d-1$ dimensional metric on the torus
and $v_{0}$ a constant. 
The non-trivial component of the current density is $J^{x}_{(0)}$, which must be a constant, and we have
\begin{align}\label{eq:j0_eq}
\frac{\gamma^{1/2}}{g_{d-1}^{1/2}Z^{(0)}}J^{x}_{(0)}= \partial_{x}w+\frac{\gamma^{1/2}a_{t}^{(0)}}{g_{d-1}^{1/2}}\,v_{0}+E_{x}\,.
\end{align}
With a little effort we can now write the 
Stokes equation \eqref{eq:V_neutral2} in the form
\begin{align}
&2\,v_{0}\,\partial_{x}\left( \gamma^{-1/2}\,\partial_{x}g_{d-1}^{-1/2}\right)-Y\,v_{0}+\frac{\gamma^{1/2}a_{t}^{(0)}}{g_{d-1}^{1/2}}J^{x}_{(0)}-\partial_{x}p=-4\pi T\,\zeta_{x}\,,\label{eq:v0_eq}
\end{align}
where we have defined
\begin{align}
&Y\equiv\frac{1}{2(\gamma g_{d-1})^{1/2}}\,\left[\left(\partial_{x} \ln g_{d-1}\right)^{2} +\partial_{x}g_{ab}\,\partial_{x}g_{cd}\,g^{ac}\,g^{bd}\right]\nn
&\qquad\qquad\qquad\qquad\qquad+\frac{1}{(\gamma\,g_{d-1})^{1/2}}\,\left(\partial_{x}\phi^{(0)}\right)^{2}+\frac{\gamma^{1/2} Z^{(0)}a_{t}^{(0)}{}^{2}}{g_{d-1}^{1/2}}\,,
\end{align}
with $g_{ab}(x)$ the metric components for $ds^2_{d-1}$.
Equations \eqref{eq:j0_eq} and \eqref{eq:v0_eq} may now be used to fix the functions $w$ and $p$.
Since these are periodic functions, we must have that the expressions for $\partial_x w$ and $\partial_x p$ have
no zero modes on the torus and this imposes constraints on $J^{x}_{(0)}$ and $v_{0}$.
A simple way to establish these constraints is take an average of the two equations. In fact doing this completely fixes $J^{x}_{(0)}$ and $v_{0}$ in terms of the sources. Indeed, if we define
\begin{align}\label{onedave}
\int \qquad\leftrightarrow \qquad  \frac{1}{L_1}\int dx^1\,,
\end{align}
where $L_1$ is the period of the lattice, we obtain
\begin{align}
J^{x}_{(0)}&=\frac{1}{X}\left(E_{x}\int Y+\,4\pi T\,\zeta_{x}\int \frac{\gamma^{1/2}a_{t}^{(0)}}{g_{d-1}^{1/2}}\right)\,,\nn
v_{0}&=\frac{1}{X}\left(4\pi T\,\zeta_{x}\,\int \,\frac{\gamma^{1/2}}{ g_{d-1}^{1/2}Z^{(0)}}+E_{x}\int \frac{\gamma^{1/2}a_{t}^{(0)}}{g_{d-1}^{1/2}}\right)\,,
\end{align}
where
\begin{align}\label{defx}
X&\equiv \left(\int \,\frac{\gamma^{1/2}}{g_{d-1}^{1/2} Z^{(0)}} \right)\left(\int Y\right)-\left(\int \frac{\gamma^{1/2}a_{t}^{(0)}}{g_{d-1}^{1/2}} \right)^{2}\,.
\end{align}
The expression for the heat current at the horizon is simply $Q^x_{(0)}=4\pi T\, v_{0}$. Now for these one-dimensional lattices the electric current and heat current densities are independent of the radial direction and so we have deduced
their values at the AdS boundary. Thus we can immediately extract the thermoelectric 
conductivities in the $x$ direction and we find
\begin{align}\label{eq:1d_DC_coeffs}
\sigma=\frac{1}{X}\int Y,\quad \alpha=\bar{\alpha}=\frac{4\pi}{X}\int \frac{\gamma^{1/2}a_{t}^{(0)}}{g_{d-1}^{1/2}},\quad \bar{\kappa}=\frac{\left(4\pi\right)^{2}T}{X}\,\int \,\frac{\gamma^{1/2}}{g_{d-1}^{1/2} Z^{(0)}}\,.
\end{align}
Observe that the final result for the conductivity is invariant under reparametrisations
of the $x$ coordinate, as it should be.

We can also write the electrical conductivity in the form
\begin{align}\label{qeqzer}
\sigma=\sigma_{Q=0}
+\frac{1}{X}\left(\int \frac{\gamma^{1/2}a_{t}^{(0)}}{g_{d-1}^{1/2}} \right)^{2}\left(\int \,\frac{\gamma^{1/2}}{g_{d-1}^{1/2} Z^{(0)}}\right)^{-1}
\end{align}
where $\sigma_{Q=0}\equiv  \sigma-T\alpha\bar\kappa^{-1}\bar\alpha$ is the conductivity when $Q=0$ (as opposed to $\zeta=0$) and is given
by
\begin{align}
\sigma_{Q=0}=\left(\int \,\frac{\gamma^{1/2}}{g_{d-1}^{1/2} Z^{(0)}}\right)^{-1}\,.
\end{align}
Notice that the second term in \eqref{qeqzer} vanishes if $a_{t}^{(0)}=0$.

We finish by indicating how to obtain some previous results. For the 
one-dimensional lattice of $D=4$ Einstein-Maxwell theory given in 
\cite{Donos:2014yya} we simply need to use the translation given by
\begin{align}
 \gamma=\Sigma(x)\,e^{B(x)},\quad ds^{2}_{1}=\Sigma(x)\,e^{-B(x)}\,dy^{2},\quad
 G^{(0)}=H_{tt}^{(0)}(x)\,,
\end{align}
in order to obtain the expressions for $\sigma,\alpha,\bar\alpha,\bar\kappa$ given in \cite{Donos:2014yya}.
Similarly for the helical lattice of pure $D=5$ gravity studied in 
\cite{Donos:2014gya} we should use
\begin{align}\label{eq:ex_helical}
 \gamma&=h^{2}_{+},\quad ds^{2}_{2}=r^{2}_{+}\,\left( e^{2\alpha_{+}}\,\omega_{2}^{2}+e^{-2\alpha_{+}}\,\omega_{3}^{2}\right)\,,
 \end{align}
where two of the three left-invariant one-forms for Bianchi VII$_0$ are given by
\begin{align}
\omega_{2}&=\cos kx\,dy-\sin k x\, dz,\quad \omega_{3}=\sin kx\,dy+\cos k x\, dz\, .
\end{align}
It is straightforward to show that $Y=4k^2\sinh^2 2\alpha_+/(r_+^2 h_+)$ and hence
recover the formula for $\kappa$ in the $x$ direction given in \cite{Donos:2014gya}.
Finally charged helical lattices in D=5 gravity with two gauge-fields coupled to a scalar were
studied in \cite{Donos:2014oha}. Setting the second gauge-field to zero, we can compare our results by setting
\begin{align}\label{eq:ex_helical}
 \gamma&=C_1,\quad ds^{2}_{2}=C_2\omega_{2}^{2}+C_3\omega_{3}^{2}\,.
 \end{align}
A short calculation shows that in the $x$ direction we have
\begin{align}
\sigma&=\frac{C_2^{1/2}C_3^{1/2}Z_0}{C_1^{1/2}}+\frac{C_2C_3}{k^2(C_2-C_3)^2}\frac{4\pi\rho^2}{s}\,,\nn
\alpha&=\bar\alpha=\frac{C_2C_3}{k^2(C_2-C_3)^2}4\pi\rho\,,\qquad
\bar\kappa=\frac{C_2C_3}{k^2(C_2-C_3)^2}4\pi sT\,,
\end{align}
where $s=4\pi (C_1C_2C_3)^{1/2}$ and $\rho=(C_1C_2C_3)^{1/2}a^{(0)}_{t}Z^{(0)}$ (see \eqref{rhotot}). The expression
for $\sigma$ agrees with \cite{Donos:2014oha} and the expressions for $\alpha,\bar\alpha$ and $\bar\kappa$ are new.

\subsection{Perturbative lattices}
We now consider the case of a periodic lattice that is constructed as a perturbative expansion about
the electrically charged AdS-RN black brane solution with a flat horizon. 
As we have noted before since everything is periodic in the spatial directions, in effect, 
we can take $\Sigma_d$ to be a torus.
If $\lambda$ is the perturbative parameter, then at the black hole horizon we will assume that we can write
\begin{align}\label{eq:pert_exp}
g_{(0)}{}_{ij}&=g\,\delta_{ij}+\lambda\,h^{(1)}_{ij}+\cdots\,,\qquad G^{(0)}=f_{(0)}+\lambda\,f_{(1)}+\cdots\,, \notag\\
{Z^{(0)}a_{t}^{(0)}}&=a+\lambda\, a_{(1)}+\cdots\,,\qquad\quad
\phi^{(0)}=\psi_{(0)}+\lambda\,\psi_{(1)}+\cdots\,,
\nn
Z^{(0)}&=z_{(0)}+\lambda\,z_{(1)}+\cdots\,,
\end{align}
with $a$, $z_{(0)}$, $\psi_{(0)}$, $f_{(0)}$ and $g$ being constant and the sub-leading terms are periodic functions of, generically, all of the spatial coordinates $x^i$. Note that the entropy density and the electric current density
on the horizon are given by
\begin{align}
s=4\pi g^{d/2},\qquad \rho_H=a g^{d/2}\,.
\end{align}
For the Ricci tensor and Christoffel symbols we have the expansions
\begin{align}
R_{(0)}{}_{ij}&=\lambda\,R^{(1)}_{ij}+\lambda^{2}\,R^{(2)}_{ij}+\cdots\,,\nn
\Gamma^i_{jk}&=\lambda \Gamma^{(1)i}_{jk}+\lambda^2 \Gamma^{(2)i}_{jk}+\cdots\,.
\end{align}

It turns out that we can solve equations \eqref{eq:v_eq2}-\eqref{eq:V_neutral2} perturbatively in $\lambda$ using the following
expansion:
\begin{align}
v^i&=\frac{1}{\lambda^{2}}\,v_{(0)}^i+\frac{1}{\lambda}\,v_{(1)}^i+v_{(2)}^i+\cdots\,,\qquad
w=\frac{1}{\lambda}\,w_{(1)}+w_{(2)}+\cdots\,,\notag\\
p&=\frac{1}{\lambda}\,p_{(1)}+p_{(2)}+\cdots\,.
\end{align}
Expanding \eqref{eq:v_eq2}-\eqref{eq:V_neutral2} in $\lambda$, we find at leading order that
\begin{align}\label{eq:v0_const}
\partial_{i}\,v_{(0)}^{i}&=0\,,\qquad
\Box\,v^{i}_{(0)}=0\,,
\end{align}
where $\Box=\delta^{ij}\partial_i\partial_j$ and we deduce that $v^{i}_{(0)}$ are just constant on the torus. We proved earlier that for the full non-linear problem, in the absence of horizon Killing vectors,
there is a unique solution to the Stokes equations.  Therefore, it must be the case that these integration constants
are fixed at higher orders in the perturbative expansion, and this will be confirmed shortly.

At next order in the expansion we find
\begin{align}\label{po}
\partial_{i}v^{i}_{(1)}+\frac{g^{-1}}{2}\,\partial_{j}h^{(1)}\,v_{(0)}^{j}&=0\,,\nn
g^{-1}z_{(0)}\,\Box w_{(1)}+v_{(0)}^{i}\,\partial_{i}\,a^{(1)}&=0\,,\nn
\Box v_{(1)}^{i}+\partial^{k}(\Gamma^{(1)}){}^{i}_{ks}v_{(0)}^{s}+R^{(1)i}{}_{j}\,v_{(0)}^{j}+a\,\partial^{i}w_{(1)}-\partial^{i}p_{(1)}&=0\,,
\end{align}
where
\begin{align}
h^{(1)}&=h^{(1)k}{}_{k}\,,\qquad
(\Gamma^{(1)}){}^{i}_{ks}=\frac{g^{-1}}{2}\left(\partial_{s}h^{(1)i}{}_{k} +\partial_{k}h^{(1)i}{}_{s}-\partial^{i}h^{(1)}{}_{ks} \right)\,,
\end{align}
and all indices are raised and lowered with $\delta$.
By considering the pressure Poisson equation \eqref{poisseq} and using $\nabla_{j}R^{j}{}_{i}=\frac{1}{2}\,\nabla_{i}R$,
we deduce that
\begin{align}\label{eq:wpp_sol}
-a\,w_{(1)}+p_{(1)}&=(\Box^{-1}\partial_{j}R^{(1)})\,v_{(0)}^{j}\,,
\end{align}
with
$w_{(1)}=-g z_{(0)}^{-1}\,(\Box^{-1}\partial_{j}a^{(1)})\,v_{(0)}^{j}$ from \eqref{po}.
Combining these results we can obtain an expression for $v_{(1)}^{i}$ in terms of $v_{(0)}^{i}$:
\begin{align}
v_{(1)}^{i}&=N^{i}_{(1)}{}_{j}\,v^{j}_{(0)}\,,
\label{eq:v1_sol}
\end{align}
with
\begin{align}
N^{i}_{(1)}{}_{j}&=-\Box^{-1}\,\left(\partial^{k}(\Gamma^{(1)}){}^{i}_{kj}+R^{(1)i}{}_{j} - \partial^{i}\,\left( \Box^{-1}\partial_{j}R^{(1)}\right)\right).\label{eq:v1_sol_2}
\end{align}
The function $\Box^{-1}f$ is defined up to a constant on a torus. The associated constant for $v_{(1)}^{i}$ will
be fixed at third order in the perturbative expansion and it does not affect the leading order DC result.

We now integrate equation \eqref{eq:V_neutral2} to find
\begin{align}\label{eq:int_eq}
\int\,g_{(0)}^{1/2}\,\nabla^{(k}v^{l)}\partial_{j}g_{(0)}{}_{kl}-&\int\, g_{(0)}^{1/2}\,{Z^{(0)}a_{t}^{(0)}}\,\partial_{j}w
+\int\, g_{(0)}^{1/2}\,\partial_{j}p
+\int\,g_{(0)}^{1/2}\,\,\partial_{i}\phi^{(0)}\partial_{j}\phi^{(0)}v^{j}\notag\\
=&\left(\int g^{1/2}_{(0)}\right)\,4\pi T\,\zeta_{j}+\left(\int\,g_{(0)}^{1/2}{Z^{(0)}a_{t}^{(0)}}\right)\,E_{j}\,,
\end{align}
where we have taken $E_i,\zeta_i$ to be constants and
$\int$ is defined to be the average over a period:
\begin{align}
\int \qquad\leftrightarrow \qquad  \frac{1}{L_1\dots L_d}\int dx^1\dots dx^d\,.
\end{align}
We next expand equation \eqref{eq:int_eq} with respect to $\lambda$ and keep the $\lambda^0$ pieces. 
Using \eqref{eq:wpp_sol}, \eqref{eq:v1_sol}, after some work we find that the left hand side can be expressed
in terms of $v^{i}_{(0)}$. Indeed we deduce that
\begin{align}
\lambda^{-2} L_{ji}v^{i}_{0}=4\pi T\,\zeta_{j}+a\,E_{j}\,,
\end{align}
where $L$ is a matrix that only depends on the background data given by 
\begin{align}
L_{ji}=&{\lambda^2}g^{-1}\int\left(\frac{g^{-1}}{2} \partial_{j}h^{(1)}_{kl}\partial_{i}h^{(1)}{}^{kl}+\partial_{j}h^{(1)}_{kl}\,\partial^{k}N^{l}{}_{i}+\frac{1}{2}\,h^{(1)}\,\partial_{j}(\Box^{-1}\partial_{i}R^{(1)})\right)\notag\\
&+\lambda^{2}\,gz_{(0)}^{-1}\,\int  a_{(1)}\,\partial_{j}\left( \Box^{-1} \partial_{i}a_{(1)}\right)+\lambda^{2}\,\int \partial_{i}\psi_{(0)}\partial_{j}\psi_{(0)}\,,
\end{align}
with $N$ as given by \eqref{eq:v1_sol_2}. Notice, in particular, that the integration constants associated with $\Box^{-1}$
drop out since the relevant terms are all covered by an extra spatial derivative.
Thus at leading order we have
\begin{align}
v^{i} &\approx\,(L^{-1})^{ij}\,\left(4\pi T\,\zeta_{j}+a\,E_{j}\right)\,,
\end{align}
and
\begin{align}
J^{i}|_H\approx \rho_H v^i\,,\qquad
Q^{i}|_H\approx  Ts v^i\,.
\end{align}

Recalling the definition of the radially independent current flux densities
given in \eqref{avecur},\eqref{avecur2}, we finally obtain the holographic current flux densities in terms of $E,\zeta$:
\begin{align}
\bar J^{i}\approx \rho v^i\,,\qquad
\bar Q^{i}\approx  Ts v^i\,,
\end{align}
where we used \eqref{rhotot}. Thus we can determine the leading order behaviour of the conductivities:
\begin{align}\label{eq:pert_DC_coeffs}
\bar\kappa=L^{-1}4\pi sT\,,\qquad \alpha=\bar\alpha=L^{-1}4\pi \rho\,,\qquad \sigma=L^{-1}\frac{4\pi \rho^2}{s}\,.
\end{align}
We observe that $\bar\kappa(\sigma T)^{-1}=s^2/\rho^2$, which corresponds to a kind of Wiedemann-Franz law.
In addition we note that the thermal conductivity at zero current flow, $\kappa\equiv \bar\kappa -T\bar\alpha\sigma^{-1}\alpha$, 
appears at a higher order in the expansion: $\kappa\sim \lambda^0$.
It is interesting to compare these results to the discussion in \cite{Mahajan:2013cja}. We similarly
find that $\sigma_{Q=0}$ also appears at a higher order in the expansion.

\subsubsection{Perturbative one-dimensional lattices}
We conclude this subsection by discussing the special case of one-dimensional
perturbative lattices and hence make contact with the results of section \ref{sec:1d_lattices}. We first notice that for an arbitrary periodic function of a single coordinate $x$ we have
\begin{align}
\Box^{-1}\partial_{x}F=c+\int_{0}^{x}dx\, F(x)-{x}\,\int \,F\,,
\end{align}
where $c$ is an arbitrary constant and in the last term $\int$ refers to the average
integral over a period in the $x$ direction, as given in \eqref{onedave}. Hence
\begin{align}
\int \,  F\,\partial_{x}\left( \Box^{-1} \partial_{x}F\right)=\int\, F^{2}-\left(\int\, F\right)^{2}\,.
\end{align}
Recalling \eqref{onedform} we next assume that
\begin{align}
h_{ij}^{(1)}\,dx^{i}dx^{j}=\delta\gamma\,dx^{2}+\delta g_{ab}\,dx^{a}dx^{b}\,,
\end{align}
where $\delta\gamma$ and $\delta g_{ab}$ only depend on $x$.
The only non-zero matrix element of the matrix $L$ is given by
\begin{align}
\lambda^{-2}\,L_{xx}=&g^{-2}\,\int \left(\frac{1}{2}\,\partial_{x}\delta g_{ab}\,\partial_{x}\delta g^{ab}+\frac{1}{2}\,\partial_{x}\delta g^{a}{}_{a}\,\partial_{x}\delta g^{b}{}_{b} \right)+\int\left(\partial_{x}\psi\right)^{2}\nn
&+gz_{(0)}^{-1}\,\left(\int a_{(1)}^{2}-\left( \int a_{(1)}\right)^{2} \right)\,,
\end{align}
where indices are raised with $\delta^{ab}$. In obtaining the above result we used that
\begin{align}
R_{xx}=-\frac{1}{2g}\,\partial_{x}^{2}\,\delta g^{a}{}_{a},\quad R_{ab}=-\frac{1}{2g}\,\partial_{x}^{2}\,\delta g_{ab}\,.
\end{align}

In the notation of section \ref{sec:1d_lattices} we have
\begin{align}
\gamma=g+\lambda\,\delta\gamma,\quad g_{ab}=g\,\delta _{ab}+\lambda\,\delta g_{ab}\,,
\end{align}
and the rest of the functions are expanded exactly as in \eqref{eq:pert_exp}. After this identification we find that $X$ as defined in \eqref{defx} takes the form
\begin{align}
X = \frac{1}{g^{d}\,z_{(0)}}\,g\,L_{xx}\,.
\end{align}
It is then straightforward to see that the leading order expansion 
in the lattice strength of the DC conductivities give in \eqref{eq:1d_DC_coeffs} 
agree with the perturbative results given in \eqref{eq:pert_DC_coeffs} when restricted to lattices that depend on one spatial coordinate only.

\section{Discussion}\label{disc}
The main results that we have obtained in previous sections apply in a more general setting as we now discuss. Specifically, we consider the following ansatz for a general class
of static solutions
\begin{align}
ds^2&=g_{tt}dt^2+ ds^2(M_{D-1}),\qquad A=a_t dt,
\end{align}
where $g_{tt},a_t,\phi$ and the metric $ds^2(M_{D-1})$ are all independent of time and are just functions of 
the coordinates $x^a$ on $M_{(D-1)}$. 

The spacetime may have various types of asymptotic boundaries, but our primary interest is when there is an $AdS$ boundary\footnote{We can also consider other holographic boundary conditions, for example, asymptotically Lifshitz, or
even asymptotically flat boundary conditions.}. In this case we can introduce a local radial coordinate and then impose the same boundary conditions as in
\eqref{eq:DC_ansatz},\eqref{asymptsol}, corresponding to the CFT living on $\Sigma_d$ deformed with various spatially dependent sources.
The spacetime
may have one or possibly more black hole horizons (examples have been discussed in \cite{Horowitz:2014gva}).
Near each black hole horizon we can again introduce a local radial coordinate and then demand that the metric has the behaviour that we gave in \eqref{nhexpbh}. Note that we do not assume that
the topology of the black hole horizons are all the same, nor do we assume that
that have the same topology as $\Sigma_d$.

We now consider the following linear perturbation
\begin{align}\label{pertansatz2}
\delta \left(ds^2\right)&=\delta g_{\mu\nu} dx^\mu dx^\nu+2t g_{tt}\zeta _adt dx^a\,,\nn
\delta A&=\delta a_\mu dx^\mu-t E_adx^a+t a_t \zeta_a dx^a\,,\nn
\delta\phi\,,
\end{align}
where the one-forms $E,\zeta$ are now defined
on $M_{D-1}$ (not just on $\Sigma_d$ as before) and are still taken to be closed.
In addition $\delta g_{\mu\nu}$, $\delta a_\mu$ and $\delta\phi$ are all independent of $t$. It is an interesting fact that at linearised order in the perturbation (and independent of any boundary conditions) we still have the key results
\begin{align}\label{gencrs}
di_k*Z(\phi)F=0,\qquad d i_k *G=0\,,
\end{align}
where $k=\partial_t$. 
Note that $k$ still satisfies the conditions \eqref{kcon}, and we have defined $G$
as before in \eqref{geedefone}. 

We now consider the boundary conditions on the perturbation. At the
holographic boundary $E,\zeta$ approach closed one-forms $E^0,\zeta^0$ on $\Sigma_d$
and we impose that these are the only sources deforming the CFT.
Similarly, for each black hole horizon the perturbation behaves in local coordinates as in \eqref{eq:nh_exp}, with
$E,\zeta$ again approaching closed one-forms
on each horizon. Using the local coordinates at each horizon
we can now impose the Hamiltonian, momentum and Gauss-law constraints
exactly as described in section \ref{consdesc} and obtain a set of generalised Stokes equations on each horizon. By solving these equations we can thus obtain currents on each horizon. Note that the precise source terms that appear in the Stokes equations on each horizon 
follows after imposing that $E,\zeta$ are closed one-forms in the bulk and that they approach $E^0,\zeta^0$ at the $AdS$ boundary.

At the $AdS$ boundary we can define the current fluxes through each $d-1$ cycle ${\cal C}_a$ on $\Sigma_d$ as in \eqref{defjay}. 
Now consider any orientable $d$-dimensional manifold in the bulk with boundary ${\cal C}_a$ at the AdS boundary and
a $d-1$ cycle at the black hole horizon (which might be disconnected). Then
using \eqref{gencrs} and Stokes's theorem, we deduce that the current fluxes are equal to the fluxes on the black hole horizon.
In turn these fluxes can be obtained by solving the Stokes fluid equations
on the black hole horizon, which only depend on the cohomology class of the sources at the horizon, which in turn only depend on the cohomology class of $E^0,\zeta^0$ on $\Sigma_d$ at the AdS boundary. Thus by
expanding the fluxes $E^0,\zeta^0$ in a basis of harmonic one-forms that are Poincar\'e dual to the ${\cal C}_a$ we have 
a procedure for obtaining the DC conductivities. 

There are a number of interesting directions to pursue. In our
analysis we assumed that the black holes have vanishing magnetic field.
In \cite{dggm} we will relax this condition and generalise the results obtained in 
\cite{Blake:2014yla,Blake:2015ina} for Q-lattices.

More generally, we have shown the DC conductivity of the boundary theory can be 
obtained by solving the generalised Stokes equations at the black hole horizon. In a narrow sense
the fluid equations are simply an auxiliary set of equations to solve this holographic problem. However,
the innate physical character of the equations (with their novel viscous terms) suggest that their
may be a deeper significance. As a first step, it would be interesting to determine whether
the full time-dependent and non-linear generalised Navier-Stokes equations at the black hole horizon can also be used to obtain exact holographic information for
the dual CFT. It is natural to expect that the time-dependent equations will be useful
in extracting the small frequency behaviour of the AC conductivity. A related point
is to use our results to develop a systematic hydrodynamic framework in the
presence of holographic lattices. Finally it would also be very interesting to obtain
some explicit lattice black hole solutions that depend on more than one of
the spatial dimensions and analyse the fluid equations. 

\section*{Acknowledgements}
We thank Tom Griffin and Luis Melgar for helpful discussions.
The work is supported by STFC grant ST/J0003533/1, EPSRC grant EP/K034456/1,
 and by the European Research Council under the European Union's Seventh Framework Programme (FP7/2007-2013), ERC Grant agreement ADG 339140. 
This research was supported in part by Perimeter Institute of Theoretical Physics. 
Research at Perimeter Institute is supported by the Government of Canada through Industry Canada and by the Province of Ontario through the Ministry of Economic Development \& Innovation.

\appendix

\section{Radial Hamiltonian formalism}\label{radham}

In this section we rewrite the equations of motion corresponding to the Lagrangian density:
\begin{align}
{\mathcal L}=&\sqrt{-g}\,\left(R-V(\phi)-\frac{Z(\phi)}{4}\,F^{2}-\frac{1}{2}\left(\partial\phi \right)^2\right)\,,
\end{align}
using a Hamiltonian decomposition with respect to the radial variable. We will
follow the notation of \cite{Wald:1984rg}, {\it mutatis mutandi}, generalising to include 
the gauge-field and the scalar field. A useful reference is \cite{Gourgoulhon:2007ue} and we note that closely related work independently appeared recently in \cite{Lindgren:2015lia}.

We consider a foliation by slices of constant $r$. We introduce the normal vector $n^\mu$, satisfying $n^\mu n_\mu=1$.
The $D$-dimensional metric $g_{\mu\nu}$ induces a $(D-1)$-dimensional Lorentzian 
metric on the 
slices of constant $r$ via $h_{\mu\nu}=g_{\mu\nu}-n_\mu n_\nu$.
The lapse and shift vectors are given by $n_\mu=N(dr)_\mu$ and $N^\mu=h^\mu{}_\nu r^\mu=r^\mu-Nn^\mu$ where
$r^\mu=(\partial_r)^\mu$. In a local coordinates system we can write
\begin{align}\label{loccoords}
ds^2=N^2 dr^2+\gamma_{ab}(dx^a+N^a dr)(dx^b+N^b dr)\,,
\end{align}
where the shift vector has components $N^\mu=(0, N^a)$ and $h_{\mu\nu}$ has components $h_{rr}=N^aN^b\gamma_{ab}$,
$h_{ra}=\gamma_{ab} N^b$ and $h_{ab}=\gamma_{ab}$.
Note also that $N_\mu=(N^bN^c\gamma_{bc},\gamma_{ab}N^a)$.

We will decompose the gauge-field components via 
\begin{align}
b_\mu=h_\mu{}^\nu A_\nu,\qquad
\Phi=-Nn^\mu A_\mu\,,
\end{align}
and hence $A_\mu=b_\mu-N^{-1}\,\Phi\, n_\mu $. In the local coordinates we have $b_r=N^a A_a$, $b_a=A_a$ and $\Phi=-A_r+N^aA_a$. 

The radial Hamiltonian formulation can be obtained by first rewriting the Lagrangian density  as follows
\begin{align}
{\mathcal L}
=&N\sqrt{-h}\,\Bigg({}^{(D-1)}R+K^{2}-K_{\mu\nu}K^{\mu\nu}-V-\frac{1}{4}Zf_{\mu\nu}f_{\rho\sigma}h^{\mu\rho}h^{\nu\sigma}
-\frac{1}{2}ZX_\mu h^{\mu\nu}X_\nu\nn
&\qquad\qquad\qquad\qquad-\frac{1}{2}h^{\mu\nu}\partial_{\mu}\phi\partial_\nu-\frac{1}{2}(n^\mu\partial_\mu\phi) (n^\nu\partial_\nu\phi)\Bigg)\,,
\end{align}
where 
we have neglected total divergences. Here
$K_{\mu\nu}=\frac{1}{2}{\mathcal L}_n h_{\mu\nu}=h_\mu{}^\rho\nabla_\rho n_\nu$ is the extrinsic curvature,
$K=g^{\mu\nu}K_{\mu\nu}$ and
\begin{align}
f_{\mu\nu}&=\partial_\mu b_\nu-\partial_\nu b_\mu\,,\nn
X_\mu&=f_{\mu\nu}n^\nu+\frac{\Phi}{N}n^\nu\nabla_\nu n_\mu -D_\mu \left(\frac{\Phi}{N}\right)\,,\nn
&=f_{\mu\nu}n^\nu-\frac{1}{N}D_\mu\Phi\,,
\end{align}
where the second expression utilises the fact that
$n^\mu\nabla_\mu n_\nu=-\frac{1}{N}D_\nu N$. This latter result follows from writing
$n_\nu=N\nabla_\nu r$ and using $\nabla_{\mu\nu} r=\nabla_{\nu\mu} r$. We also recall
that $D_\mu$ is the Levi-Civita connection associated with the metric $h$ and, for example,
$D_\mu\Phi=h_\mu{}^\nu\nabla_\nu\Phi$.

With $\dot h_{\mu\nu}={\mathcal L}_r h_{\mu\nu}$, 
$\dot b_{\mu}={\mathcal L}_r b_{\mu}$, $\dot\phi={\cal L}_r\phi$ we can show that
\begin{align}
\dot h_{\mu\nu}&=2NK_{\mu\nu}+D_\mu N_\nu+D_\nu N_\mu\,,\nn
\dot b_\mu&=r^\rho f_{\rho\mu}+\nabla_\mu(b_\rho N^\rho)\,,\nn
&=Nn^\rho f_{\rho\mu}+N^\rho f_{\rho\mu}+\nabla_\mu(b_\rho N^\rho)\,,\nn
\dot\phi&=Nn^\mu\partial_\mu \phi+N^\mu\partial_\mu\phi\,.
\end{align}
The corresponding conjugate momenta are then given by
\begin{align}\label{momdefns}
\pi^{\mu\nu}&=\frac{\delta {\mathcal L}}{\delta \dot{h}_{\mu\nu}}=-\sqrt{-h}\,\left( K^{\mu\nu}-K\,h^{\mu\nu}\right)\,,\notag\\
\pi^\mu&=\frac{\delta {\mathcal L}}{\delta \dot{b}_{\mu}}=\sqrt{-h}h^{\mu\rho} ZX_\rho\,,\nn
&\qquad\quad =\sqrt{-h} Z F^{\mu\rho}n_\rho\,,\nn
\pi_\phi&=\frac{\delta {\mathcal L}}{\delta \dot\phi}=-\frac{\sqrt{-h}}{N}\left(\dot\phi-N^\nu\partial_\nu\phi\right)\,,\nn
&\qquad\quad=-\sqrt{-h}n^\nu\partial_\nu\phi\,,
\end{align}
where the second expressions for $\pi^\mu$ and $\pi_\phi$, which are not written in the canonical variables, are
useful.

The Hamiltonian density, defined
as $\mathcal{H}= \pi^{\mu\nu}\dot h_{\mu\nu}+\pi^\mu \dot b_\mu -{\mathcal L}$,
can be written as a sum of constraints
\begin{align}
\mathcal{H}&=
N\,H+N_{\mu}\,H^{\mu}+\Phi\,C \,,
\end{align}
with
\begin{align}
H=&\,-(-h)^{-1/2}\,\left(\pi_{\mu\nu}\pi^{\mu\nu}-\frac{1}{D-2}\,\pi^{2} \right)-\sqrt{-h}\,\left({}^{(D-1)}R-V\right)\nn
&-\frac{1}{2}\,(-h)^{-1/2}\,Z^{-1}h_{\mu\nu}\,\pi^{\mu}\,\pi^{\nu}+\frac{1}{4}\,\sqrt{-h}\,Zf_{\mu\nu}\,f_{\rho\sigma}\,h^{\mu\rho}\,h^{\nu\sigma}\nn
&-\frac{1}{2}(-h)^{-1/2}\pi^2_\phi+\frac{1}{2}\sqrt{-h} h^{\rho\sigma}\partial_\rho\phi\partial_\sigma\phi
\label{eq:H_constraint}\,,\\
H^{\nu}=&-2\sqrt{-h}\,D_{\mu}\left((-h)^{-1/2}\pi^{\mu\nu}\right)+h^{\nu\sigma}f_{\sigma\rho} \pi^\rho\nn
&-h^{\nu\sigma}a_\sigma\sqrt{-h} D_\rho\left( (-h)^{-1/2}\, \pi^\rho\right)
+h^{\nu\sigma}\partial_\sigma\phi\pi_\phi \,,\label{eq:Hn_constraint}\\
C=&\sqrt{-h}\,D_{\mu}\left( (-h)^{-1/2}\,\pi^{\mu}\right)\,,\label{eq:G_constraint}
\end{align}
where $\pi=\pi^\mu{}_\mu$ and we have ignored total divergences\footnote{In passing that we note that in the local coordinates \eqref{loccoords}
we have $\Phi-N^\mu a_\mu=-A_r$ and one sees that using explicit coordinates one could uses $A_r$ as a Lagrange multiplier instead of $\Phi$.}.

The equations of motion are given by 
\begin{align}\label{keyeqs}
\dot{h}_{\mu\nu}=&- 2N(-h)^{-1/2}\,\left(\pi_{\mu\nu}-\frac{1}{d}\,\pi\,h_{\mu\nu} \right)+2\,D_{\left(\mu\right.}N_{\left.\nu\right)}\,,\nn
\dot{\pi}^{\mu\nu}=&-N\,\sqrt{-h}\,\left({}^{(d+1)}R^{\mu\nu} - \frac{1}{2}{}^{(d+1)}R\,h^{\mu\nu}+\frac{1}{2}V\,h^{\mu\nu}\right)\notag\\
&-\frac{1}{2}N\,(-h)^{-1/2}\,h^{\mu\nu}\,\left(\pi_{\gamma\delta}\pi^{\gamma\delta}-\frac{1}{d}\,\pi^{2} \right)
+2N(-h)^{-1/2}\,\left( \pi^{\mu\gamma}\,\pi^{\nu}{}_{\gamma}-\frac{1}{d}\,\pi\,\pi^{\mu\nu}\right)\notag\\
&+\sqrt{-h}\,\left( D^{\mu}D^{\nu}N-h^{\mu\nu}\,D^{\gamma}D_{\gamma}N\right)
\notag\\
&+\frac{1}{2}N(-h)^{-1/2}Z^{-1}\left(\pi^\mu\pi^\nu-\frac{1}{2}h^{\mu\nu}(h_{\rho\sigma}\pi^\rho\pi^\sigma)\right)\nn
&+\frac{1}{2}N\sqrt{-h}Z\left(h^{\mu\lambda}h^{\nu\gamma}h^{\rho\sigma}f_{\lambda\rho}f_{\gamma\sigma}
-\frac{1}{4}h^{\mu\nu}(h^{\rho\sigma}h^{\gamma\delta}f_{\rho\gamma}f_{\sigma\delta})\right)
\nn
&-\frac{1}{4}N(-h)^{-1/2}\pi_\phi^2 h^{\mu\nu}+\frac{1}{2}N\sqrt{-h}\left(h^{\mu\rho}h^{\nu\sigma}\partial_\rho\phi\partial_\sigma\phi-\frac{1}{2}(h^{\rho\sigma}\partial_\rho\phi\partial_\sigma\phi) h^{\mu\nu}\right)\nn
&+\sqrt{-h}\,D_{\gamma}\left((-h)^{-1/2} N^{\gamma} \pi^{\mu\nu}\right)-2\,\pi^{\gamma\,\left(\mu \right.}D_{\gamma}N^{\left. \nu\right)}
\end{align}
as well as
\begin{align}
\dot{b}_{\mu}=&D_{\mu}(N^\nu b_\nu)-N\,(-h)^{-1/2}\,\pi_{\mu}+N^{\nu}\,f_{\nu\mu}-D_{\mu}\Phi\,,\nn
\dot{\pi}^{\mu}=& \sqrt{-h}\,D_{\sigma}\left(NZ\,h^{\sigma\rho}h^{\mu\delta}f_{\rho\delta}\right)
+2\sqrt{-h}\,D_{\sigma}\left((-h)^{-1/2}\,N^{\left[ \sigma\right.}\pi^{\left.\mu\right]} \right)\,,
\end{align}
where we have dropped pieces proportional to the constraints in $\dot\pi^{\mu\nu}$ and $\dot\pi^\mu$.
We also have 
\begin{align}
\dot\phi=&-(-h)^{-1/2} N\pi_\phi+N^\nu\partial_\nu\phi\,,\nn
\dot\pi_\phi=&\sqrt{-h}D_\mu(\sqrt{-h}\pi_\phi N^\mu)+\sqrt{-h}D_\mu(Nh^{\mu\nu}D_\nu\phi)\nn
&-N\Bigg(\sqrt{-h}V'+\frac{1}{2}\,(-h)^{-1/2}\,Z^{-2}Z'h_{\mu\nu}\,\pi^{\mu}\,\pi^{\nu}+\frac{1}{4}\,\sqrt{-h}\,Z'f_{\mu\nu}\,f_{\rho\sigma}\,h^{\mu\rho}\,h^{\nu\sigma}\Bigg)\,.
\end{align}

\section{Generalised Stokes equations from the constraints}\label{navst}
In this appendix we show how the Stokes equations \eqref{eq:v_eq2}-\eqref{eq:V_neutral2}
arise from the constraint equations \eqref{eq:H_constraint}-\eqref{eq:G_constraint} in a radial decomposition. 
More precisely we will examine the constraints for the perturbed metric, at linearised order, focussing on the leading terms
of an expansion at the black hole horizon. In other words, we evaluate the constraints on a
hypersurface of constant $r$ and then take the limit $r\to 0$.

We begin by noting that for the perturbed metric the unit normal vector has components
\begin{align}
n^{i}&=-U^{1/2}F^{-1/2}\,g_{d}^{ij}\,\delta g_{rj},\qquad n^{t}=G^{-1} \left(FU\right)^{-1/2}\,\delta g_{tr},\nn
n^{r}&=U^{1/2}F^{-1/2}\,\left( 1-\frac{U}{2F}\,\delta g_{rr}\right)\,.
\end{align}
Furthermore, the corresponding shift and lapse functions are given by
\begin{align}
N^{j}&=g_{d}^{ij}\,\delta g_{ri},\qquad N^{t}=-\frac{1}{GU}\,\delta g_{rt}\,,\nn
N&=F^{1/2}U^{-1/2}\,\left( 1+\frac{1}{2}\,\frac{U}{F}\,\delta g_{rr}\right)\,,
\end{align}
The components of the extrinsic curvature take the form
\begin{align}\label{genkay}
K^{tt}=&\frac{1}{2}\,G^{-2}U^{-3/2}F^{-1/2}\,\left(-\partial_{r}\left(GU\right)+\frac{1}{2}\,\frac{U}{F}\,\partial_{r}\left(GU\right) \,\delta g_{rr}\right)\nn
&+\frac{1}{2}\,G^{-2}U^{-3/2}F^{-1/2}\,\left(\left(GU\right)^{2} \partial_{r}\left( \frac{\delta g_{tt}}{\left(GU\right)^{2}}\right)+\partial_{j}(GU)\,N^{j}\right)\,,\nn
K^{ti}=&\frac{1}{2}\,U^{1/2}F^{-1/2}\,\left(-\partial_{r}\left( \frac{1}{GU}\,g_{d}^{ij}\,\left(\delta g_{tj}
-tGU\zeta_j\right)\right)+g_{d}^{ij}\,\partial_{j}\left(\frac{1}{GU}\,\delta g_{rt} \right) \right)\,,\nn
K^{ij}=&-U^{1/2}F^{-1/2}\,\nabla^{(i}N^{j)}
+\frac{1}{2}\,U^{1/2}F^{-1/2}\,\left( \frac{U}{2F}\partial_{r}g_{d}^{ij}\,\delta g_{rr}-\partial_{r}g_{d}^{ij}+g^{ik}_dg^{jl}_d\partial_{r}\delta g_{kl}\right)\nn
&-U^{1/2}F^{-1/2}\,g^{l(i}_dg^{j)m}_dg^{kn}_d\delta g_{mn}\,\partial_{r}g_{d}{}_{kl}\,.
\end{align}
where here $\nabla$ is the covariant derivative compatible with the $d$-dimensional metric $g_{d}{}_{ij}$. 

Expanding the extrinsic curvature
close to the horizon we find
\begin{align}\label{kayathor}
K^{tt}&\rightarrow -\frac{1}{2}\frac{1}{(4\pi T)^{1/2}}\,\frac{1}{r^{3/2}}\frac{1}{G^{(0)}{}^{3/2}}\,\left(1+\frac{\delta g^{(0)}_{tt}}{G^{(0)}}-\frac{1}{2}\frac{\delta g^{(0)}_{rr}}{G^{(0)}}+\frac{1}{4\pi T}v^{i}\,\partial_{i}\ln G^{(0)} \right)\,,\nn
K^{ti}&\rightarrow -\frac{1}{2}\,\frac{1}{(4\pi T)^{1/2}}\frac{1}{r^{3/2}}\,\frac{1}{G^{(0)}{}^{3/2}}\,v^{i}\,,\nn
K^{ij}&\rightarrow \frac{1}{(4\pi T)^{1/2}}\frac{1}{r^{1/2}}\,\frac{1}{G^{(0)}{}^{1/2}}\,\nabla^{(i}v^{j)}\,,\nn
K&\rightarrow \frac{1}{2}\,\frac{(4\pi T)^{1/2}}{G^{(0)}{}^{1/2}r^{1/2}}\,\left( 1-\frac{1}{2}\frac{\delta g^{(0)}_{rr}}{G^{(0)}}+\frac{1}{4\pi T}\,\nabla_{i}v^{i}+\frac{1}{4\pi T}\,v^{i}\nabla_{i}\ln G^{(0)}\right)\,,
\end{align}
and we also note that
\begin{align}
K^{t}{}_{i}\rightarrow \frac{1}{2}\,\frac{(4\pi T)^{1/2}}{G^{(0)}{}^{1/2}r^{1/2}}\,\zeta_{i}\,t\,.
\end{align}
We now consider the following quantity which appears in the momentum constraint \eqref{eq:Hn_constraint}
\begin{align}
W_{\nu}=&D_{\mu}\left((-h)^{-1/2}\pi^{\mu}{}_{\nu}\right)=-D_{\mu}K^{\mu}{}_{\nu}+D_{\nu}K\nn
=&-(-h)^{-1/2}\,\partial_{\mu}\left(\sqrt{-h}\,K^{\mu}{}_{\nu}\right)+\frac{1}{2}\,\partial_{\nu}h_{\kappa\lambda}K^{\kappa\lambda}+\partial_{\nu}K\,.
\end{align}
Expanding at the horizon we find the following individual components
\begin{align}\label{wstuf}
W_{t}&\rightarrow -\frac{1}{2}\frac{(4\pi T)^{1/2}}{G^{(0)}{}^{1/2}}\frac{1}{r^{1/2}}\,\nabla_{i}v^{i}\,,\nn
W_{i}&\rightarrow \frac{1}{G^{(0)}{}^{1/2}}\frac{1}{(4\pi T)^{1/2}}\frac{1}{r^{1/2}}\,\left(-\nabla^{j}\nabla_{(j}v_{i)} - 2\pi T\,\zeta_{j}+\frac{1}{2}\,\nabla_{i}p\right)
\end{align}
where
\begin{align}\label{pexpression}
p=-2 \pi T\frac{1}{G^{(0)}}\,\left(\delta g_{tt}^{(0)}+g_{rr}^{(0)}\right)-\delta g_{it}^{(0)}g^{ij}_{(0)}\nabla_{j}\,\ln G^{(0)}\,.
\end{align}
Notice that after imposing the boundary condition constraints \eqref{eq:nh_constr} this definition of $p$ is identical to the definition of pressure
given in \eqref{deffquants}.
Another quantity that enters the constraints is the momentum of the scalar field. At leading order in $r$ we have 
\begin{align}\label{scmom}
\pi_{\phi}\rightarrow -\sqrt{g_{(0)}}\,v^{i}\partial_{i}\phi\,.
\end{align}
We now turn to the gauge field. From the second expression in \eqref{momdefns}
we have
\begin{align}\label{pinh}
\pi^{\mu}=\sqrt{-h}\,Z\,F^{\mu\lambda}\,n_{\lambda}=\sqrt{-g}\,Z\,F^{\mu r}=J^{\mu}\,.
\end{align}
After expanding near the horizon we also find
\begin{align}\label{fstuff}
f_{t\mu}\pi^{\mu}&=f_{ti}J^{i}\rightarrow 0\,,\nn
f_{i\mu}\pi^{\mu}&=f_{it}J^{t}+f_{ij}J^{j}\rightarrow \left(\partial_{i}w+E_{i} \right)J^{t}=g_{(0)}^{1/2}\,{Z^{(0)}\,a_{t}^{(0)}}\,\left(\partial_{i}w+E_{i} \right)\,.
\end{align}

We now now consider the constraint equations.
Substituting \eqref{pinh} into the Gauss constraint \eqref{eq:G_constraint}, $\partial_\mu\pi^\mu=0$, we obtain the current continuity equation 
$\partial_{\mu} J^{\mu}=0$. When evaluated at the horizon this leads to $\partial_i J^i_{(0)}=0$. We next consider the momentum constraints
 \eqref{eq:Hn_constraint} with lowered indices. Using the above results we find that the $t$ component gives $\nabla_i v^i=0$ while the $i$
 component gives the Stokes equation \eqref{eq:V_neutral2}. 
 
Finally, examining the Hamiltonian constraint \eqref{eq:H_constraint}, 
we find that the leading order expansion at the horizon for the terms involving the linearised perturbation implies $\nabla_i v^i=0$, and hence gives no further conditions. To see this we consider $H$ in \eqref{eq:H_constraint}
as a sum of six terms and it is convenient to divide by $\sqrt{-h}$. 
Using \eqref{scmom} and \eqref{pinh} we immediately see that the third and
fifth terms vanish at linearised order. It turns out that the leading order power of $r$ that appears is $r^{-1}$. We can show that the sixth term and, with a bit more effort,
the fourth terms are of order $r^{0}$. Next we consider the second term. The potential
term is clearly of order $r^0$. After examining the leading terms in the Christoffel
symbols we can also show that the Ricci scalar term is also of this order.
Finally, we need to examine the first term. To do so it is convenient to note
that using \eqref{kayathor} we have
\begin{align}
(-h)^{-1/2}\pi^{tt} &\to -\frac{1}{2}\frac{1}{(4\pi T)^{3/2}G^{(0)}{}^{3/2}}\frac{1}{ r^{3/2}} \nabla_i v^i \, , \nn
(-h)^{-1/2}\pi^{ij} &\to\frac{1}{2}\frac{1}{(4\pi T)^{1/2}G^{(0)}{}^{1/2}}\frac{1}{r^{1/2}}\Big(-2\nabla^{(i}v^{j)} - 4 \pi T \delta g_{kl}^{(0)}g_{(0)}^{ik}g_{(0)}^{jl} \notag \\
&+4\pi T g_{(0)}^{ij} (1- \frac{1}{2G^{(0)}}\delta g_{rr}^{(0)}+\frac{1}{4 \pi T}\nabla_i v^i+\frac{1}{4 \pi T}v^i\nabla_i \ln G^{(0)})\Big)\, ,
\end{align}
Continuing to evaluate the first term we are eventually led to the result that the leading term in the Hamiltonian constraint can be written
\begin{align}
(-h)^{-1/2}H &\to
\frac{1}{2 r G^{(0)}} \nabla_i v^i.
\end{align}

\section{Holographic currents}\label{holcurr}

On-shell we have
\begin{align}
\delta S=\int d^3x\sqrt{-h}\left[\tfrac{1}{2} \left(r^{-(D+1)}t^{\mu\nu}\right)\delta h_{\mu\nu}
+ \left(r^{1-D}j^\mu\right)\delta A_\mu\right]\,.
\end{align}
where $t^{\mu\nu}$ and $j^\mu$ are the radially independent, holographic stress tensor and current, respectively.
After substituting the time dependent sources given in \eqref{pertansatz}, \eqref{mandn} we find
\begin{align}
\delta S=\int d^3 x\bar G^{1/2}\sqrt{\bar g_d}\left[  (\bar Gt^{ti}-\mu j^{i})(-t\hsce_i )+j^{i}(-tE_i)\right]\,.
\end{align}
We thus see that $-tE_i$ is a source for the operator density $\bar G^{1/2}\sqrt{\bar g_d} j^i$ and
$-t\hsce_i$ is a source for the operator density $\bar G^{1/2}\sqrt{\bar g_d} (\bar Gt^{ti}-\mu j^{i})$.

It is possible to show that the expectation values of these holographic tensor densities are given by
\begin{align}\label{denten}
\bar G^{1/2}\sqrt{\bar g_d} j^i&=J^i|_\infty\,,\nn
\bar G^{1/2}\sqrt{\bar g_d} (\bar Gt^{ti}-\mu j^{i})&= Q^i|_\infty- t \bar G^{3/2} \sqrt{\bar g_d} t^{ij} \zeta_{j} \, .
\end{align}
Thus $J^i|_\infty$ and $Q^i|_\infty$ are the time-independent parts of the expectation values of the
vector and tensor densities.  To establish the first equation in \eqref{denten} is straightforward. The second
is a little more involved. Firstly, 
from the expression for $Q^i$ given in \eqref{defqueue} and using \eqref{genkay}
we can show that at linearised order we have
\begin{align}
Q^i=F(GU)^{3/2}\sqrt{g_d}\left(-2K^{ti}+2K^{ij}\frac{1}{GU}\left( \delta g_{tj}
-tGU\zeta_j  \right)\right)-a_tJ^i\,.
\end{align}
Recall that if we write 
\begin{align}
\tilde t^{\mu\nu}=-2K^{\mu\nu}+X h^{\mu\nu}+ Y^{\mu\nu}\,,
\end{align}
where $X=2 K +f(\phi)+\dots$ and $Y$ corresponds to additional terms arising from the counterterms,
then we obtain the stress tensor if we evaluate $\tilde t^{\mu\nu}$ at the AdS boundary. 
Observing that at linearised order we have
$g_{tt}h^{ti}=h^{ij}\left( \delta g_{tj}
-tGU\zeta_j  \right)$
we can therefore write
\begin{align}
Q^i=(GU)^{3/2}\sqrt{g_d}\left((\tilde t^{ti}-Y^{ti})-(\tilde t^{ij}-Y^{ij})\frac{1}{GU}\left( \delta g_{tj}
-tGU\zeta_j  \right)\right)-a_tJ^i\,.
\end{align}
We next want to take a limit as $r\to\infty$. If the combination of $Y^{ti}$ and $Y^{ij}$ that appear make sub-leading contributions, then we have
\begin{align}
Q^i|_\infty=\lim_{r\to\infty}\left[r^{D+1}\bar G^{3/2}\sqrt{g_d}\left(\tilde t^{ti}+\tilde t^{ij}t\zeta_j  \right)-a_tJ^i\right]
\end{align}
and we recover \eqref{denten} after using that $r^{D+1} \tilde t^{\mu\nu}=t^{\mu\nu}$.
We have explicitly checked for particular cases, eg $D=4$ with $Y^{\mu\nu}\sim  R^{(3)\mu\nu}$ that
this does indeed occur. It would be interesting to find a universal argument that this is always true.

The time dependent piece on the right hand side of \eqref{denten} 
is associated with the static susceptibility for the
heat current two point function, as explained in appendix C of \cite{Donos:2014cya}.
It can also be understood by noting that if we start with the background black hole geometries, with in particular $t^{ti}=j^i=0$, 
then the time independent linear perturbation that is generated
by the coordinate transformation $t\to t+\zeta_i x^i$, induces the 
transformation $t^{ti}\to -\zeta_i t^{ij}$. Promoting this perturbation to 
one that is linear in time leads to the time dependence as in \eqref{denten}.

\section{Alternative derivation of the Stokes equations}\label{otherder}
We discussed in section \ref{curhor} how the generalised Stokes system
of equations given in \eqref{eq:v_eq2}-\eqref{eq:V_neutral2}
can also be obtained from the equations 
\eqref{jandfeoms}, \eqref{Qeqns} for $J^i$ and $Q^i$. Indeed
evaluating the first of the two equations in each of 
\eqref{jandfeoms}, \eqref{Qeqns}
at the black hole horizon we immediately obtain 
$\partial_iJ^i_{(0)}=\partial_iQ^i_{(0)}=0$. These comprise two
of the three Stokes equations, given in \eqref{eq:v_eq2}, \eqref{eq:w_eq2}. 
The third Stokes equation, given in \eqref{eq:V_neutral2},
can be obtained from the second equation of \eqref{Qeqns}.

To obtain it
we consider the pieces of $Q^i$ that are linear in $r$ obtaining
\begin{align}\label{lefths}
&\sqrt{g_{(0)}}\left[-G^{(0)}g^{ij}_{({(0)})}  \partial_j\left(4\pi T\frac{\delta g^{(0)}_{rt}}{G^{(0)}}\right)      
-\delta g^{(0)}_{jt}M^{ij}
-4\pi T G^{(0)}\zeta_j\right] -a^{(0)}_t J^i_{(0)}\,,
\end{align}
where we have defined the matrix
\begin{align}\label{eq:M_def}
M^{ij}=\,g^{ij}_{(0)}\,\left[4\pi T\,\left( \frac{3G^{(1)}}{2G^{(0)}}-\frac{F^{(1)}}{2G^{(0)}}\right) +2U^{(1)} \right]+4\pi T\,g_{(0)}^{-1/2} \left(\sqrt{g} g^{ij}\right)^{(1)}\,.
\end{align}
Notice that this matrix 
depends on next to leading order terms in the expansion at the black hole horizon.
Equation \eqref{Qeqns} then implies that \eqref{lefths} should equal
\begin{align}
-\partial_j\left[G^{(0)}{}^{2}\sqrt{g_{(0)}}g^{jk}_{(0)}g^{il}_{(0)}\left(\partial_k\left(\frac{\delta g_{lt}^{(0)}}{G^{(0)}}\right)-k\leftrightarrow l\right)\right]\,.
\end{align}
To obtain an equation at the black hole horizon, we need to be able to express the matrix $M$ in terms of leading order horizon data. After a long calculation, which we outline below, using the equations of motion for the background black hole
we can show the key result
\begin{align}\label{eq:M_tensortext}
M^{ij}=&-2G^{(0)}\,{}^{(d)}R^{ij}+4\,G^{(0)}{}^{1/2}\,\nabla^{i}\nabla^{j} G^{(0)}{}^{1/2}\,-\,g_{(0)}^{ij}\,\Box G^{(0)}\nn&+Z\,{G^{(0)}}a_{t}^{(0)}{}^{2}\,g_{(0)}^{ij}
+G^{(0)}g_{(0)}^{ik}g_{(0)}^{jl}\partial_k\phi^{(0)}\partial_l\phi^{(0)}\,,
\end{align}
and this leads to the final Stokes equation in \eqref{eq:V_neutral2}.

We use the radial Hamiltonian presentation of the equations of motion
for the background black hole solutions \eqref{eq:DC_ansatz}. 
The unit normal vector
is $n=U^{1/2}\,F^{-1/2}\,\partial_{r}$. 
The lapse function is given by $N=U^{-1/2}F^{1/2}$ and the shift vector vanishes,
$N^{\mu}=0$.
The non-vanishing components of the extrinsic curvature are given by
\begin{align}
K_{tt}&=-\frac{1}{2}\,U^{1/2}\,F^{-1/2}\,\partial_{r}\left( U G\right)\,,\nn
K_{ij}&=\frac{1}{2}\,U^{1/2}\,F^{-1/2}\,\partial_{r}g_{ij}\,,
\end{align}
and hence the non-vanishing components of the conjugate momentum are
given by
\begin{align}
\pi^{tt}&=-\left(FG\right)^{-1/2}\,\partial_{r}\,g^{1/2}\,,\nn
\pi^{ij}&=-\frac{1}{2}\,U\,\left(G\,g\right)^{1/2}\,F^{-1/2}\,g^{ik}\,g^{jl}\,\partial_{r}\,g_{kl}+U^{1/2}\,F^{-1/2}\,g^{ij}\,\partial_{r}\,\left( UGg\right)^{1/2}\,,\nn
\pi^{t}&=(FG)^{-1/2}\,g^{1/2}Z\,\partial_{r}a_{t}\,,\nn
\pi_\phi&=-G^{1/2}F^{1/2}g^{1/2}\dot\phi \,.
\end{align}
We also have
\begin{align}
\pi=h_{\mu\nu}\pi^{\mu\nu}=d\,U^{1/2}\,F^{-1/2}\,\partial_{r}\left( U\,G\,g\right)^{1/2}.
\end{align}

It is convenient to rewrite the equation of motion for $\pi^{\mu\nu}$ given in 
\eqref{keyeqs} in the form
\begin{align}\label{eq:pi_eom}
&\dot{\pi}^{\mu\nu}-2N(-h)^{-1/2}\,\left( \pi^{\mu\gamma}\,\pi^{\nu}{}_{\gamma}-\frac{1}{d}\,\pi\,\pi^{\mu\nu}\right)=\nn
&-N\,\sqrt{-h}\,\left({}^{(d+1)}R^{\mu\nu} - {}^{(d+1)}R\,h^{\mu\nu}+V\,h^{\mu\nu}\right)
+\sqrt{-h}\,\left( D^{\mu}D^{\nu}N-h^{\mu\nu}\,D^{\gamma}D_{\gamma}N\right)\notag\\
&+\frac{1}{2}NZ\sqrt{-h}\left(h^{\mu\lambda}h^{\nu\gamma}h^{\rho\sigma}f_{\lambda\rho}f_{\gamma\sigma}
-\frac{1}{2}h^{\mu\nu}(h^{\rho\sigma}h^{\gamma\delta}f_{\rho\gamma}f_{\sigma\delta})\right)
+\frac{1}{2}N(-h)^{-1/2}Z^{-1}\pi^\mu\pi^\nu\nn
&+\frac{1}{2}N\sqrt{-h}\left(h^{\mu\rho}h^{\nu\sigma}\partial_\rho\phi\partial_\sigma\phi-h^{\rho\sigma}\partial_\rho\phi\partial_\sigma\phi\,h^{\mu\nu}\right)\,,
\end{align}
where we used the fact that $N^{\mu}=0$ as well as the constraint $H=0$ with $H$ as in \eqref{eq:H_constraint}.

We now wish to plug in the background expansions \eqref{nhexpbh} in the equations of motion \eqref{eq:pi_eom}. Taking the $r\rightarrow 0$ limit of the left hand side yields
\begin{align}
\dot{\pi}^{tt}-2N(-h)^{-1/2}\,\left( \pi^{t\gamma}\,\pi^{t}{}_{\gamma}-\frac{1}{d}\,\pi\,\pi^{tt}\right)&\rightarrow- \frac{1}{r}\,\frac{1}{G^{(0)}}\,\left(\sqrt{g}\right)^{(1)}\,,\nn
\dot{\pi}^{i j}-2N(-h)^{-1/2}\,\left( \pi^{i \gamma}\,\pi^{j}{}_{\gamma}-\frac{1}{d}\,\pi\,\pi^{i j}\right)&\rightarrow \frac{ \sqrt{g_{(0)}}}{2}\,M^{ij}+4\pi T\,g_{(0)}^{ij}\,\left(\sqrt{g}\right)^{(1)}\,,
\end{align}
with $M^{ij}$ as defined in equation \eqref{eq:M_def}. 
For the right hand side of \eqref{eq:pi_eom}
we find the following leading order behaviour as $r\rightarrow 0$:
\begin{align}\label{ttandij}
(tt)\rightarrow & -\frac{1}{r}\frac{1}{4\pi T}\,g_{(0)}^{1/2}\,\left({}^{(d+1)}R^{tt}_{(0)}\,G^{(0)} \,4\pi T r+{}^{(d+1)}R_{(0)}-V_{(0)} \right) \notag\\
&+\frac{1}{r}\frac{1}{4\pi T}\,g_{(0)}^{1/2}\,G^{(0)}{}^{1/2}\,\left( -\frac{1}{2}\frac{1}{G^{(0)}{}^{2}} g_{(0)}^{ij}\,\partial_{i}G^{(0)}\,\partial_{j}G^{(0)}{}^{1/2}+\frac{1}{G^{(0)}}\,D^{\gamma}D_{\gamma} G^{(0)}{}^{1/2}\right)\notag\\
&+\frac{1}{r}\frac{1}{4\pi T}\frac{1}{2\,G^{(0)}{}^{2}}\,g_{(0)}^{1/2}\,Za_{t}^{(0)}{}^{2}+
\frac{1}{4\pi r T}\frac{1}{2}g_{(0)}^{1/2}g_{(0)}^{ij}\partial_i\phi^{(0)}\partial_j\phi^{(0)}\,,
\end{align}
\begin{align}
(ij)\rightarrow &- G^{(0)}g_{(0)}^{1/2}\,\left( {}^{d+1}R_{(0)}^{ij}-{}^{d+1}R_{(0)}\,g_{(0)}^{ij}+V_{(0)}\,g_{(0)}^{ij}\right)\notag\\
&+ \sqrt{G^{(0)}\,g^{(0)}}\,\left(D^{i}D^{j}\,G^{(0)}{}^{1/2}-g_{(0)}^{ij}\,D_{\gamma}D^{\gamma}\,G^{(0)}{}^{1/2} \right)\nn
&+\frac{1}{2}G^{(0)}g_{(0)}^{1/2}\left( g_{(0)}^{ik}g_{(0)}^{jl}\partial_k\phi^{(0)}\partial_l\phi^{(0)}
-g_{(0)}^{kl}\partial_k\phi^{(0)}\partial_l\phi^{(0)}\,g_{(0)}^{ij}
\right)\,.
\end{align}

We now decompose the $d+1$ dimensional Ricci tensor scalar via:
\begin{align}
{}^{(d+1)}R_{ij}=&{}^{(d)}R_{ij}-\frac{1}{2}\,\nabla_{i}\left(\frac{\nabla_{j}G}{G} \right)-\frac{1}{4}\,G^{-2}\,\nabla_{i}G\,\nabla_{j}G\,,\nn
{}^{(d+1)}R_{tt}=&\left(\frac{1}{2}\,\nabla_{i}\left(\frac{\nabla^{i}G}{G} \right)+\frac{1}{4}\,G^{-2}\,\nabla_{i}G\,\nabla^{i}G \right)\,GU\,,\nn
{}^{(d+1)}R=&{}^{(d)}R-\nabla_{i}\left(\frac{\nabla^{i}G}{G} \right)-\frac{1}{2}\,G^{-2}\,\nabla_{i}G\,\nabla^{i}G\,.
\end{align}
We also have
\begin{align}
D^{\gamma}D_{\gamma}G^{(0)}{}^{1/2}=\frac{1}{2}\,G^{(0)}{}^{-1}\,g_{(0)}^{ij}\,\partial_{i}G^{(0)}\,\partial_{j}G^{(0)}{}^{1/2}+\nabla^{i}\nabla_{i}G^{(0)}{}^{1/2}\,,
\end{align}
where $\nabla$ is the covariant derivative with respect to the $d$ dimensional horizon metric. Putting these ingredients together we finally obtain the expression for
$M^{ij}$ given in \eqref{eq:M_tensortext}.


\begin{thebibliography}{10}

\bibitem{Policastro:2001yc}
G.~Policastro, D.~T. Son, and A.~O. Starinets, ``{The Shear viscosity of
  strongly coupled N=4 supersymmetric Yang-Mills plasma},''
  \href{http://dx.doi.org/10.1103/PhysRevLett.87.081601}{{\em Phys.Rev.Lett.}
  {\bfseries 87} (2001) 081601},
\href{http://arxiv.org/abs/hep-th/0104066}{{\ttfamily arXiv:hep-th/0104066
  [hep-th]}}.

\bibitem{Kovtun:2004de}
P.~Kovtun, D.~T. Son, and A.~O. Starinets, ``{Viscosity in strongly interacting
  quantum field theories from black hole physics},''
  \href{http://dx.doi.org/10.1103/PhysRevLett.94.111601}{{\em Phys.Rev.Lett.}
  {\bfseries 94} (2005) 111601},
\href{http://arxiv.org/abs/hep-th/0405231}{{\ttfamily arXiv:hep-th/0405231
  [hep-th]}}.

\bibitem{Donos:2015gia}
A.~Donos and J.~P. Gauntlett, ``{Navier-Stokes on Black Hole Horizons and DC
  Thermoelectric Conductivity},''
\href{http://arxiv.org/abs/1506.01360}{{\ttfamily arXiv:1506.01360 [hep-th]}}.

\bibitem{dggm}
A.~Donos, J.~P. Gauntlett, T.~Griffin, and L.~Melgar,
``{DC conductivity of holographic magnetised matter},"
{\em {To appear}} .

\bibitem{Vegh:2013sk}
D.~Vegh, ``{Holography without translational symmetry},''
\href{http://arxiv.org/abs/1301.0537}{{\ttfamily arXiv:1301.0537 [hep-th]}}.

\bibitem{Blake:2013owa}
M.~Blake, D.~Tong, and D.~Vegh, ``{Holographic Lattices Give the Graviton a
  Mass},'' \href{http://dx.doi.org/10.1103/PhysRevLett.112.071602}{{\em
  Phys.Rev.Lett.} {\bfseries 112} (2014) 071602},
\href{http://arxiv.org/abs/1310.3832}{{\ttfamily arXiv:1310.3832 [hep-th]}}.

\bibitem{Davison:2013txa}
R.~A. Davison, K.~Schalm, and J.~Zaanen, ``{Holographic duality and the
  resistivity of strange metals},''
  \href{http://dx.doi.org/10.1103/PhysRevB.89.245116}{{\em Phys.Rev.}
  {\bfseries B89} (2014) 245116},
\href{http://arxiv.org/abs/1311.2451}{{\ttfamily arXiv:1311.2451 [hep-th]}}.

\bibitem{Amoretti:2014mma}
A.~Amoretti, A.~Braggio, N.~Maggiore, N.~Magnoli, and D.~Musso, ``{Analytic dc
  thermoelectric conductivities in holography with massive gravitons},''
  \href{http://dx.doi.org/10.1103/PhysRevD.91.025002}{{\em Phys.Rev.}
  {\bfseries D91} no.~2, (2015) 025002},
\href{http://arxiv.org/abs/1407.0306}{{\ttfamily arXiv:1407.0306 [hep-th]}}.

\bibitem{Baggioli:2014roa}
M.~Baggioli and O.~Pujolas, ``{Electron-Phonon Interactions, Metal-Insulator
  Transitions, and Holographic Massive Gravity},''
  \href{http://dx.doi.org/10.1103/PhysRevLett.114.251602}{{\em Phys. Rev.
  Lett.} {\bfseries 114} no.~25, (2015) 251602},
\href{http://arxiv.org/abs/1411.1003}{{\ttfamily arXiv:1411.1003 [hep-th]}}.

\bibitem{Amoretti:2015gna}
A.~Amoretti and D.~Musso, ``{Universal formulae for thermoelectric transport
  with magnetic field and disorder},''
\href{http://arxiv.org/abs/1502.02631}{{\ttfamily arXiv:1502.02631 [hep-th]}}.

\bibitem{Karch:2007pd}
A.~Karch and A.~O'Bannon, ``{Metallic AdS/CFT},''
  \href{http://dx.doi.org/10.1088/1126-6708/2007/09/024}{{\em JHEP} {\bfseries
  0709} (2007) 024},
\href{http://arxiv.org/abs/0705.3870}{{\ttfamily arXiv:0705.3870 [hep-th]}}.

\bibitem{Hartnoll:2009ns}
S.~A. Hartnoll, J.~Polchinski, E.~Silverstein, and D.~Tong, ``{Towards strange
  metallic holography},'' \href{http://dx.doi.org/10.1007/JHEP04(2010)120}{{\em
  JHEP} {\bfseries 1004} (2010) 120},
\href{http://arxiv.org/abs/0912.1061}{{\ttfamily arXiv:0912.1061 [hep-th]}}.

\bibitem{Horowitz:2012ky}
G.~T. Horowitz, J.~E. Santos, and D.~Tong, ``{Optical Conductivity with
  Holographic Lattices},''
  \href{http://dx.doi.org/10.1007/JHEP07(2012)168}{{\em JHEP} {\bfseries 1207}
  (2012) 168},
\href{http://arxiv.org/abs/1204.0519}{{\ttfamily arXiv:1204.0519 [hep-th]}}.

\bibitem{Horowitz:2012gs}
G.~T. Horowitz, J.~E. Santos, and D.~Tong, ``{Further Evidence for
  Lattice-Induced Scaling},''
  \href{http://dx.doi.org/10.1007/JHEP11(2012)102}{{\em JHEP} {\bfseries 1211}
  (2012) 102},
\href{http://arxiv.org/abs/1209.1098}{{\ttfamily arXiv:1209.1098 [hep-th]}}.

\bibitem{Chesler:2013qla}
P.~Chesler, A.~Lucas, and S.~Sachdev, ``{Conformal field theories in a periodic
  potential: results from holography and field theory},''
  \href{http://dx.doi.org/10.1103/PhysRevD.89.026005}{{\em Phys.Rev.}
  {\bfseries D89} (2014) 026005},
\href{http://arxiv.org/abs/1308.0329}{{\ttfamily arXiv:1308.0329 [hep-th]}}.

\bibitem{Ling:2013nxa}
Y.~Ling, C.~Niu, J.-P. Wu, and Z.-Y. Xian, ``{Holographic Lattice in
  Einstein-Maxwell-Dilaton Gravity},''
  \href{http://dx.doi.org/10.1007/JHEP11(2013)006}{{\em JHEP} {\bfseries 1311}
  (2013) 006},
\href{http://arxiv.org/abs/1309.4580}{{\ttfamily arXiv:1309.4580 [hep-th]}}.

\bibitem{Balasubramanian:2013yqa}
K.~Balasubramanian and C.~P. Herzog, ``{Losing Forward Momentum
  Holographically},''
  \href{http://dx.doi.org/10.1088/0264-9381/31/12/125010}{{\em
  Class.Quant.Grav.} {\bfseries 31} (2014) 125010},
\href{http://arxiv.org/abs/1312.4953}{{\ttfamily arXiv:1312.4953 [hep-th]}}.

\bibitem{Donos:2014yya}
A.~Donos and J.~P. Gauntlett, ``{The thermoelectric properties of inhomogeneous
  holographic lattices},''
  \href{http://dx.doi.org/10.1007/JHEP01(2015)035}{{\em JHEP} {\bfseries 1501}
  (2015) 035},
\href{http://arxiv.org/abs/1409.6875}{{\ttfamily arXiv:1409.6875 [hep-th]}}.

\bibitem{Rangamani:2015hka}
M.~Rangamani, M.~Rozali, and D.~Smyth, ``{Spatial Modulation and Conductivities
  in Effective Holographic Theories},''
  \href{http://dx.doi.org/10.1007/JHEP07(2015)024}{{\em JHEP} {\bfseries 07}
  (2015) 024},
\href{http://arxiv.org/abs/1505.05171}{{\ttfamily arXiv:1505.05171 [hep-th]}}.

\bibitem{Donos:2013eha}
A.~Donos and J.~P. Gauntlett, ``{Holographic Q-lattices},''
  \href{http://dx.doi.org/10.1007/JHEP04(2014)040}{{\em JHEP} {\bfseries 1404}
  (2014) 040},
\href{http://arxiv.org/abs/1311.3292}{{\ttfamily arXiv:1311.3292 [hep-th]}}.

\bibitem{Donos:2014uba}
A.~Donos and J.~P. Gauntlett, ``{Novel metals and insulators from
  holography},'' \href{http://dx.doi.org/10.1007/JHEP06(2014)007}{{\em JHEP}
  {\bfseries 1406} (2014) 007},
\href{http://arxiv.org/abs/1401.5077}{{\ttfamily arXiv:1401.5077 [hep-th]}}.

\bibitem{Azeyanagi:2009pr}
T.~Azeyanagi, W.~Li, and T.~Takayanagi, ``{On String Theory Duals of
  Lifshitz-like Fixed Points},''
  \href{http://dx.doi.org/10.1088/1126-6708/2009/06/084}{{\em JHEP} {\bfseries
  0906} (2009) 084},
\href{http://arxiv.org/abs/0905.0688}{{\ttfamily arXiv:0905.0688 [hep-th]}}.

\bibitem{Mateos:2011ix}
D.~Mateos and D.~Trancanelli, ``{The anisotropic N=4 super Yang-Mills plasma
  and its instabilities},''
  \href{http://dx.doi.org/10.1103/PhysRevLett.107.101601}{{\em Phys.Rev.Lett.}
  {\bfseries 107} (2011) 101601},
\href{http://arxiv.org/abs/1105.3472}{{\ttfamily arXiv:1105.3472 [hep-th]}}.

\bibitem{Mateos:2011tv}
D.~Mateos and D.~Trancanelli, ``{Thermodynamics and Instabilities of a Strongly
  Coupled Anisotropic Plasma},''
  \href{http://dx.doi.org/10.1007/JHEP07(2011)054}{{\em JHEP} {\bfseries 1107}
  (2011) 054},
\href{http://arxiv.org/abs/1106.1637}{{\ttfamily arXiv:1106.1637 [hep-th]}}.

\bibitem{Andrade:2013gsa}
T.~Andrade and B.~Withers, ``{A simple holographic model of momentum
  relaxation},'' \href{http://dx.doi.org/10.1007/JHEP05(2014)101}{{\em JHEP}
  {\bfseries 1405} (2014) 101},
\href{http://arxiv.org/abs/1311.5157}{{\ttfamily arXiv:1311.5157 [hep-th]}}.

\bibitem{Cheng:2014qia}
L.~Cheng, X.-H. Ge, and S.-J. Sin, ``{Anisotropic plasma at finite $U(1)$
  chemical potential},'' \href{http://dx.doi.org/10.1007/JHEP07(2014)083}{{\em
  JHEP} {\bfseries 1407} (2014) 083},
\href{http://arxiv.org/abs/1404.5027}{{\ttfamily arXiv:1404.5027 [hep-th]}}.

\bibitem{Jain:2014vka}
S.~Jain, N.~Kundu, K.~Sen, A.~Sinha, and S.~P. Trivedi, ``{A Strongly Coupled
  Anisotropic Fluid From Dilaton Driven Holography},''
  \href{http://dx.doi.org/10.1007/JHEP01(2015)005}{{\em JHEP} {\bfseries 1501}
  (2015) 005},
\href{http://arxiv.org/abs/1406.4874}{{\ttfamily arXiv:1406.4874 [hep-th]}}.

\bibitem{Donos:2012js}
A.~Donos and S.~A. Hartnoll, ``{Interaction-driven localization in
  holography},'' \href{http://dx.doi.org/10.1038/nphys2701}{{\em Nature Phys.}
  {\bfseries 9} (2013) 649--655},
\href{http://arxiv.org/abs/1212.2998}{{\ttfamily arXiv:1212.2998}}.

\bibitem{Donos:2014gya}
A.~Donos, J.~P. Gauntlett, and C.~Pantelidou, ``{Conformal field theories in
  $d=4$ with a helical twist},''
  \href{http://dx.doi.org/10.1103/PhysRevD.91.066003}{{\em Phys.Rev.}
  {\bfseries D91} (2015) 066003},
\href{http://arxiv.org/abs/1412.3446}{{\ttfamily arXiv:1412.3446 [hep-th]}}.

\bibitem{Donos:2014oha}
A.~Donos, B.~Gout\'eraux, and E.~Kiritsis, ``{Holographic Metals and Insulators
  with Helical Symmetry},''
  \href{http://dx.doi.org/10.1007/JHEP09(2014)038}{{\em JHEP} {\bfseries 1409}
  (2014) 038},
\href{http://arxiv.org/abs/1406.6351}{{\ttfamily arXiv:1406.6351 [hep-th]}}.

\bibitem{Erdmenger:2015qqa}
J.~Erdmenger, B.~Herwerth, S.~Klug, R.~Meyer, and K.~Schalm, ``{S-Wave
  Superconductivity in Anisotropic Holographic Insulators},''
  \href{http://dx.doi.org/10.1007/JHEP05(2015)094}{{\em JHEP} {\bfseries 1505}
  (2015) 094},
\href{http://arxiv.org/abs/1501.07615}{{\ttfamily arXiv:1501.07615 [hep-th]}}.

\bibitem{Gouteraux:2014hca}
B.~Gout\'eraux, ``{Charge transport in holography with momentum dissipation},''
  \href{http://dx.doi.org/10.1007/JHEP04(2014)181}{{\em JHEP} {\bfseries 1404}
  (2014) 181},
\href{http://arxiv.org/abs/1401.5436}{{\ttfamily arXiv:1401.5436 [hep-th]}}.

\bibitem{Ling:2014laa}
Y.~Ling, P.~Liu, C.~Niu, J.-P. Wu, and Z.-Y. Xian, ``{Holographic
  Superconductor on Q-lattice},''
  \href{http://dx.doi.org/10.1007/JHEP02(2015)059}{{\em JHEP} {\bfseries 1502}
  (2015) 059},
\href{http://arxiv.org/abs/1410.6761}{{\ttfamily arXiv:1410.6761 [hep-th]}}.

\bibitem{Andrade:2014xca}
T.~Andrade and S.~A. Gentle, ``{Relaxed superconductors},''
  \href{http://dx.doi.org/10.1007/JHEP06(2015)140}{{\em JHEP} {\bfseries 06}
  (2015) 140},
\href{http://arxiv.org/abs/1412.6521}{{\ttfamily arXiv:1412.6521 [hep-th]}}.

\bibitem{Iqbal:2008by}
N.~Iqbal and H.~Liu, ``{Universality of the hydrodynamic limit in AdS/CFT and
  the membrane paradigm},''
  \href{http://dx.doi.org/10.1103/PhysRevD.79.025023}{{\em Phys.Rev.}
  {\bfseries D79} (2009) 025023},
\href{http://arxiv.org/abs/0809.3808}{{\ttfamily arXiv:0809.3808 [hep-th]}}.

\bibitem{Donos:2014cya}
A.~Donos and J.~P. Gauntlett, ``{Thermoelectric DC conductivities from black
  hole horizons},'' \href{http://dx.doi.org/10.1007/JHEP11(2014)081}{{\em JHEP}
  {\bfseries 1411} (2014) 081},
\href{http://arxiv.org/abs/1406.4742}{{\ttfamily arXiv:1406.4742 [hep-th]}}.

\bibitem{Bhattacharyya:2008jc}
S.~Bhattacharyya, V.~E. Hubeny, S.~Minwalla, and M.~Rangamani, ``{Nonlinear
  Fluid Dynamics from Gravity},''
  \href{http://dx.doi.org/10.1088/1126-6708/2008/02/045}{{\em JHEP} {\bfseries
  0802} (2008) 045},
\href{http://arxiv.org/abs/0712.2456}{{\ttfamily arXiv:0712.2456 [hep-th]}}.

\bibitem{Bhattacharyya:2008kq}
S.~Bhattacharyya, S.~Minwalla, and S.~R. Wadia, ``{The Incompressible
  Non-Relativistic Navier-Stokes Equation from Gravity},''
  \href{http://dx.doi.org/10.1088/1126-6708/2009/08/059}{{\em JHEP} {\bfseries
  0908} (2009) 059},
\href{http://arxiv.org/abs/0810.1545}{{\ttfamily arXiv:0810.1545 [hep-th]}}.

\bibitem{Fouxon:2008tb}
I.~Fouxon and Y.~Oz, ``{Conformal Field Theory as Microscopic Dynamics of
  Incompressible Euler and Navier-Stokes Equations},''
  \href{http://dx.doi.org/10.1103/PhysRevLett.101.261602}{{\em Phys.Rev.Lett.}
  {\bfseries 101} (2008) 261602},
\href{http://arxiv.org/abs/0809.4512}{{\ttfamily arXiv:0809.4512 [hep-th]}}.

\bibitem{Price:1986yy}
R.~Price and K.~Thorne, ``{Membrane Viewpoint on Black Holes: Properties and
  Evolution of the Stretched Horizon},''
\href{http://dx.doi.org/10.1103/PhysRevD.33.915}{{\em Phys.Rev.} {\bfseries
  D33} (1986) 915--941}.

\bibitem{Bredberg:2011jq}
I.~Bredberg, C.~Keeler, V.~Lysov, and A.~Strominger, ``{From Navier-Stokes To
  Einstein},'' \href{http://dx.doi.org/10.1007/JHEP07(2012)146}{{\em JHEP}
  {\bfseries 1207} (2012) 146},
\href{http://arxiv.org/abs/1101.2451}{{\ttfamily arXiv:1101.2451 [hep-th]}}.

\bibitem{Compere:2011dx}
G.~Compere, P.~McFadden, K.~Skenderis, and M.~Taylor, ``{The Holographic fluid
  dual to vacuum Einstein gravity},''
  \href{http://dx.doi.org/10.1007/JHEP07(2011)050}{{\em JHEP} {\bfseries 1107}
  (2011) 050},
\href{http://arxiv.org/abs/1103.3022}{{\ttfamily arXiv:1103.3022 [hep-th]}}.

\bibitem{Huang:2011he}
T.-Z. Huang, Y.~Ling, W.-J. Pan, Y.~Tian, and X.-N. Wu, ``{From Petrov-Einstein
  to Navier-Stokes in Spatially Curved Spacetime},''
  \href{http://dx.doi.org/10.1007/JHEP10(2011)079}{{\em JHEP} {\bfseries 1110}
  (2011) 079},
\href{http://arxiv.org/abs/1107.1464}{{\ttfamily arXiv:1107.1464 [gr-qc]}}.

\bibitem{Lysov:2013jsa}
V.~Lysov, ``{On the Magnetohydrodynamics/Gravity Correspondence},''
\href{http://arxiv.org/abs/1310.4181}{{\ttfamily arXiv:1310.4181 [hep-th]}}.

\bibitem{Hartnoll:2007ih}
S.~A. Hartnoll, P.~K. Kovtun, M.~Muller, and S.~Sachdev, ``{Theory of the
  Nernst effect near quantum phase transitions in condensed matter, and in
  dyonic black holes},''
  \href{http://dx.doi.org/10.1103/PhysRevB.76.144502}{{\em Phys. Rev.}
  {\bfseries B76} (2007) 144502},
\href{http://arxiv.org/abs/0706.3215}{{\ttfamily arXiv:0706.3215
  [cond-mat.str-el]}}.

\bibitem{Davison:2014lua}
R.~A. Davison and B.~Gout\'eraux, ``{Momentum dissipation and effective
  theories of coherent and incoherent transport},''
  \href{http://dx.doi.org/10.1007/JHEP01(2015)039}{{\em JHEP} {\bfseries 1501}
  (2015) 039},
\href{http://arxiv.org/abs/1411.1062}{{\ttfamily arXiv:1411.1062 [hep-th]}}.

\bibitem{Blake:2015epa}
M.~Blake, ``{Momentum relaxation from the fluid/gravity correspondence},''
\href{http://arxiv.org/abs/1505.06992}{{\ttfamily arXiv:1505.06992 [hep-th]}}.

\bibitem{Lucas:2015lna}
A.~Lucas, ``{Hydrodynamic transport in strongly coupled disordered quantum
  field theories},''
\href{http://arxiv.org/abs/1506.02662}{{\ttfamily arXiv:1506.02662 [hep-th]}}.

\bibitem{Davison:2015bea}
R.~A. Davison and B.~Gout\'eraux, ``{Dissecting holographic conductivities},''
\href{http://arxiv.org/abs/1505.05092}{{\ttfamily arXiv:1505.05092 [hep-th]}}.

\bibitem{Kiritsis:2015oxa}
E.~Kiritsis and J.~Ren, ``{On Holographic Insulators and Supersolids},''
\href{http://arxiv.org/abs/1503.03481}{{\ttfamily arXiv:1503.03481 [hep-th]}}.

\bibitem{Jain:2015txa}
S.~Jain, R.~Samanta, and S.~P. Trivedi, ``{The Shear Viscosity in Anisotropic
  Phases},''
\href{http://arxiv.org/abs/1506.01899}{{\ttfamily arXiv:1506.01899 [hep-th]}}.

\bibitem{Blake:2015ina}
M.~Blake, A.~Donos, and N.~Lohitsiri, ``{Magnetothermoelectric Response from
  Holography},'' \href{http://dx.doi.org/10.1007/JHEP08(2015)124}{{\em JHEP}
  {\bfseries 08} (2015) 124},
\href{http://arxiv.org/abs/1502.03789}{{\ttfamily arXiv:1502.03789 [hep-th]}}.

\bibitem{Gauntlett:1998fz}
J.~P. Gauntlett, R.~C. Myers, and P.~K. Townsend, ``{Black holes of D = 5
  supergravity},'' \href{http://dx.doi.org/10.1088/0264-9381/16/1/001}{{\em
  Class.Quant.Grav.} {\bfseries 16} (1999) 1--21},
\href{http://arxiv.org/abs/hep-th/9810204}{{\ttfamily arXiv:hep-th/9810204
  [hep-th]}}.

\bibitem{Horowitz:2014gva}
G.~T. Horowitz, N.~Iqbal, J.~E. Santos, and B.~Way, ``{Hovering Black Holes
  from Charged Defects},''
  \href{http://dx.doi.org/10.1088/0264-9381/32/10/105001}{{\em
  Class.Quant.Grav.} {\bfseries 32} no.~10, (2015) 105001},
\href{http://arxiv.org/abs/1412.1830}{{\ttfamily arXiv:1412.1830 [hep-th]}}.

\bibitem{Mahajan:2013cja}
R.~Mahajan, M.~Barkeshli, and S.~A. Hartnoll, ``{Non-Fermi liquids and the
  Wiedemann-Franz law},''
  \href{http://dx.doi.org/10.1103/PhysRevB.88.125107}{{\em Phys.Rev.}
  {\bfseries B88} (2013) 125107},
\href{http://arxiv.org/abs/1304.4249}{{\ttfamily arXiv:1304.4249
  [cond-mat.str-el]}}.

\bibitem{Blake:2014yla}
M.~Blake and A.~Donos, ``{Quantum Critical Transport and the Hall Angle},''
  \href{http://dx.doi.org/10.1103/PhysRevLett.114.021601}{{\em Phys. Rev.
  Lett.} {\bfseries 114} no.~2, (2015) 021601},
\href{http://arxiv.org/abs/1406.1659}{{\ttfamily arXiv:1406.1659 [hep-th]}}.

\bibitem{Wald:1984rg}
R.~M. Wald, {\em General Relativity}.
\newblock The University of Chicago Press, 1984.

\bibitem{Gourgoulhon:2007ue}
E.~Gourgoulhon, ``{3+1 formalism and bases of numerical relativity},''
\href{http://arxiv.org/abs/gr-qc/0703035}{{\ttfamily arXiv:gr-qc/0703035
  [GR-QC]}}.

\bibitem{Lindgren:2015lia}
J.~Lindgren, I.~Papadimitriou, A.~Taliotis, and J.~Vanhoof, ``{Holographic Hall
  conductivities from dyonic backgrounds},''
\href{http://arxiv.org/abs/1505.04131}{{\ttfamily arXiv:1505.04131 [hep-th]}}.

\end{thebibliography}

\providecommand{\href}[2]{#2}\begingroup\raggedright\endgroup

\end{document}